\documentclass[10pt,twocolumn,letterpaper]{article}
\usepackage[accsupp]{axessibility}
\usepackage{cvpr} 
\usepackage[dvipsnames]{xcolor}
\definecolor{cvprblue}{rgb}{0.21,0.49,0.74}
\usepackage[pagebackref,breaklinks,colorlinks,citecolor=cvprblue]{hyperref}

\usepackage{xspace}
\usepackage{anyfontsize}
\usepackage[dvipsnames]{xcolor}

\usepackage{amsmath}
\usepackage{amssymb}
\usepackage{amsthm}
\usepackage{amsfonts}       
\usepackage{url}            
\usepackage{booktabs}       
\usepackage{nicefrac}       
\usepackage{microtype}     
\usepackage[ruled,vlined]{algorithm2e}
\usepackage{threeparttable}
\usepackage{makecell}
\usepackage{booktabs}
\usepackage{caption}
\usepackage{wrapfig}
\usepackage{tabularx}
\usepackage{adjustbox}
\usepackage{bm}
\usepackage{tablefootnote}
\usepackage{multirow}
\usepackage{pifont}
\usepackage{array}
\usepackage{subcaption}
\usepackage{booktabs}
\usepackage{varwidth}
\usepackage{nicematrix,tikz}

\NiceMatrixOptions
  {
    custom-line = 
     {
       letter = : ,
       command = dashedline , 
       ccommand = cdashedline ,
       tikz = dashed
     }
  }
  
\newcommand{\Tref}[1]{Table~\ref{#1}}
\newcommand{\eref}[1]{Eq.~\eqref{#1}}
\newcommand{\Eref}[1]{Equation~\eqref{#1}}
\newcommand{\fref}[1]{Fig.~\ref{#1}}
\newcommand{\Fref}[1]{Figure~\ref{#1}}
\newcommand{\Sref}[1]{Section~\ref{#1}}

\newcounter{todos}
\AtEndDocument{\ifnum\value{todos}>0 \PackageWarning{TODOS}{There are \arabic{todos} todos left in this paper! Fix them before submitting the paper!} \fi}

\newcommand{\red}[1]{{\color{red}#1}}
\newcommand{\blue}[1]{{\color{blue}#1}}

\newcommand{\V}[1]{\ensuremath{\mathbf{#1}}} 

\makeatletter
\DeclareRobustCommand\onedot{\futurelet\@let@token\@onedot}
\def\@onedot{\ifx\@let@token.\else.\null\fi\xspace}

\def\eg{{\it e.g}\onedot} 
\def\ie{{\it i.e}\onedot} 
 
\def\etc{{\it etc}\onedot} 
\def\wrt{w.r.t\onedot} 
\def\etal{{\it et al}\onedot}
\makeatother

\def\paperID{5292} 
\def\confName{CVPR}
\def\confYear{2025}

\title{Noise Modeling in One Hour: Minimizing Preparation Efforts for\\Self-supervised Low-Light RAW Image Denoising}
\author{Feiran Li\\
Sony Research\\
\and
Haiyang Jiang\thanks{Work done during an internship in Sony Research.}\\
Tokyo University\\
\and
Daisuke Iso\\
Sony Research\\
}

\begin{document}
\maketitle

\begin{abstract}

Noise synthesis is a promising solution for addressing the data shortage problem in data-driven low-light RAW image denoising. However, accurate noise synthesis methods often necessitate labor-intensive calibration and profiling procedures during preparation, preventing them from landing to practice at scale. This work introduces a practically simple noise synthesis pipeline based on detailed analyses of noise properties and extensive justification of widespread techniques. Compared to other approaches, our proposed pipeline eliminates the cumbersome system gain calibration and signal-independent noise profiling steps, reducing the preparation time for noise synthesis from days to hours. Meanwhile, our method exhibits strong denoising performance, showing an up to $0.54\mathrm{dB}$ PSNR improvement over the current state-of-the-art noise synthesis technique. Code is released at \url{https://github.com/SonyResearch/raw_image_denoising}

\end{abstract}    
\section{Introduction}\label{sec:intro}
Noise, as a key factor influencing image quality, becomes more pronounced in low-light conditions due to the reduced signal-to-noise ratio. Compared to standard RGB images~\cite{zhang2017beyond, zamir2022restormer, zhang2018ffdnet}, denoising in the unprocessed RAW domain shows great potential as it retains the primitive noise characteristics and preserves more information owing to the higher bit-depth. Although learning-based approaches trained with real-world noisy-clean image pairs have made significant progress in RAW image denoising~\cite{feng2023learnability, zamir2022learning, guo2019toward, liu2019learning}, their practicality is severely limited by the need for collecting paired data. As a result, self-supervised denoising methods based on noise synthesis have gained increasing attention recently~\cite{abdelhamed2019noise, monakhova2022dancing, wei2021physics, cao2023physics, wang2020practical}.

\begin{figure}[t]
    \centering
    \includegraphics[width=\linewidth]{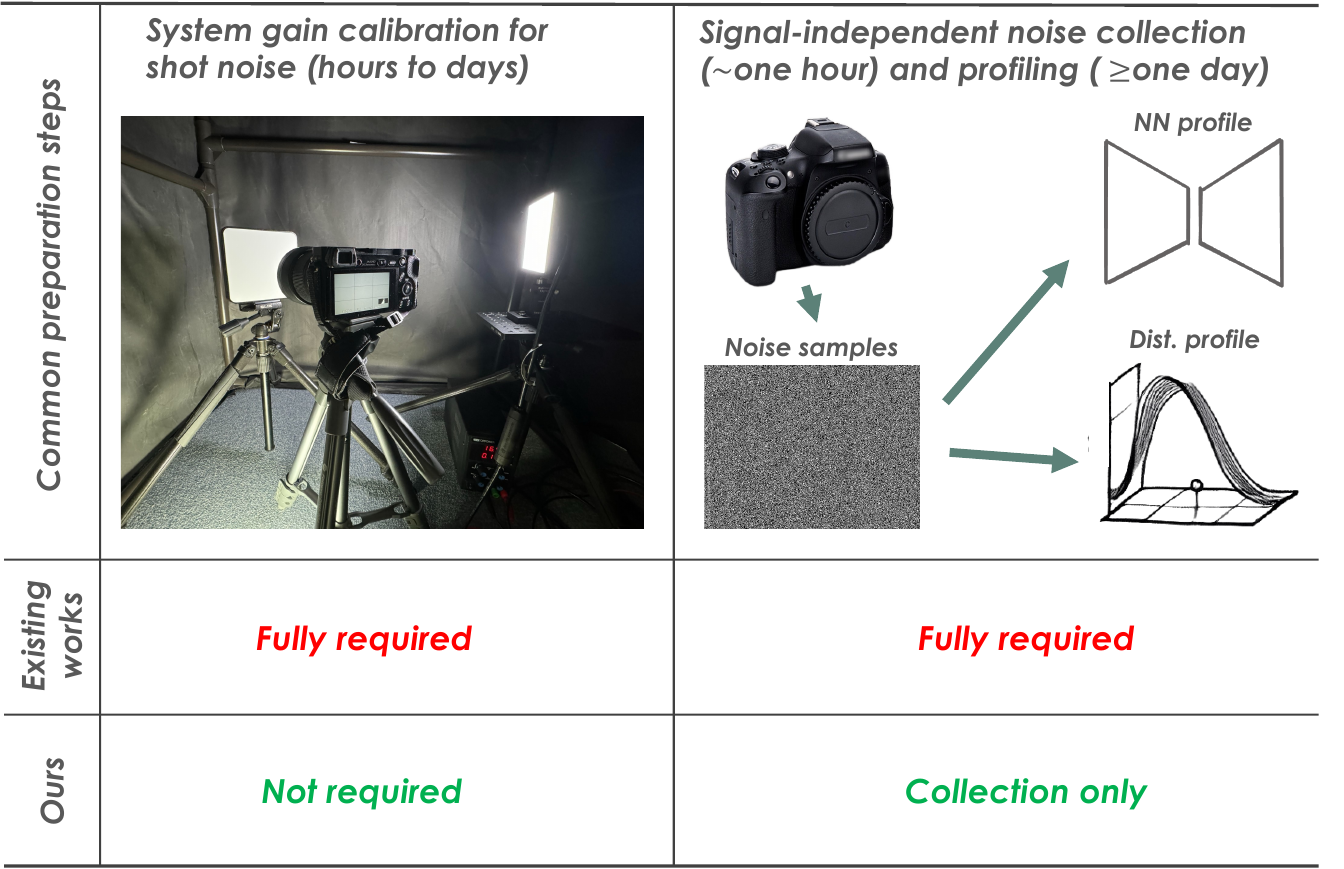}
    \caption{Comparing preparation efforts for noise synthesis. While existing methods typically spend hours to days on preparation, our pipeline requires collecting dark frames only, which is a fully automatic process and finishable within even one hour.}
    \label{fig:teaser}
\end{figure}

Precise noise generation is typically achieved through sensor-specific modeling, encompassing two steps: signal-dependent and signal-independent noise synthesis. However, these steps often require extensive human intervention and involve complex procedures. As illustrated in \fref{fig:teaser}, synthesizing signal-dependent noise demands system gain calibration under controlled lighting conditions, costing up to days for hardware setup (\eg, protocol camera boards or irregular-shaped devices like endoscopies may require customizing mounting modules). Meanwhile, signal-independent noise synthesis leads to noise profiling, necessitating careful justification of statistical distributions~\cite{wang2020practical, wei2021physics}, and even the training of intermediate profiling neural networks~\cite{feng2023physics, monakhova2022dancing, cao2023physics, zhang2023towards}. Consequently, deploying self-supervised RAW image denoising at scale remains resource-intensive.

In this work, we seek to develop a noise synthesis pipeline that, while retaining effective denoising performance, is as simple as possible for practical deployment. To achieve this, we analyze in detail the properties of photon shot noise and demonstrate that it can be adequately synthesized via hypothesizing quantum efficiencies. We also carefully justify the typical efforts spent on signal-independent noise synthesis and propose to omit any unnecessary steps therein, such as explicit noise profiling and bit-depth expansion. We summarize our key contributions as follows:
\begin{itemize}
    \item We propose a novel hypothesizing-based shot noise synthesis method that bypasses the laborious system gain calibration process. We also provide a thorough analysis \wrt signal-independent noise synthesis, highlighting the inherent limitations of parametric noise profiling and the redundancy of bit-depth expansion on dark frames.
    \item Aggregating the above, we introduce a practically simple yet effective RAW image noise synthesis pipeline that requires dark frame collection only, reducing the preparation efforts from days to hours while delivering impressive denoising results.
    \item We conduct extensive experiments to show the validity of our proposed method, and comprehensive ablation studies for an in-depth understanding. 
\end{itemize}
To the best of our knowledge, our proposal is the first approach that eliminates all the expensive calibration and profiling steps required for sensor-specific noise synthesis. The only necessary preparation effort lies in dark frame collection, which, for modern cameras, is a fully automated process finishable within even one hour. 
\section{Related works}\label{sec:related_work}
This section briefly reviews general denoising algorithms that require no noisy-clean image pairs, and recent advances in RAW image denoising that leverage physical priors.

\subsection{General image denoising without paired data}
Conventional image denoising algorithms rely on manually designed image priors. For example, numerous methods~\cite{traditional_denoise1, traditional_denoise5, traditional_denoise6, traditional_denoise7} have explored leveraging the inherent sparsity of image details to improve denoising performance. Total variation~\cite{traditional_denoise3, traditional_denoise4}, based on the observation that noise often results in high-frequency changes, is a popular regularization term for smoothing out noise. Gu~\etal~\cite{traditional_denoise2} propose exploiting the non-local self-similarity of image. Despite their long-standing use, these methods generally struggle to perform effectively on images with heavy noise, such as those captured in low-light conditions.

Learning-based methods have shown superior denoising performance. Lacking noisy-clean image pairs, most methods have focused on synthesizing photorealistic noisy images from the clean counterparts. For example, Zhu~\etal~\cite{zhu2016noise} employ Gaussian mixture models for noise modeling. Zhang~\etal~\cite{zhang2023practical} introduce a second-order degradation mixing strategy to robustify the denoising networks \wrt real-world noise. Many recent approaches incorporate neural networks for noise synthesis. For example, the flow-based and generative network-based algorithms~\cite{abdelhamed2019noise, kousha2022modeling, tran2020gan, ca_gan} aim to synthesize noise via transforming Gaussian random variables into samples that follow the underlying true noise distribution. Despite their relatively better performance, such methods are difficult to deploy in practice due to the complexity and instability introduced by training the intermediate noise synthesis networks. There also exist some works~\cite{moran2020noisier2noise, batson2019noise2self, wang2023noise2info} that explore training denoising networks with noisy images only by assuming certain types of noise (\eg, white Gaussian). However, such methods face significant domain gaps in real-world applications due to their over-simplified assumption \wrt noise distribution.

\subsection{RAW image denoising with physical priors}
A popular trend of RAW image denoising focuses on integrating physical priors into learning-based approaches, offering better noise representation in challenging scenarios. Specifically, noise in camera sensors can be categorized into signal-dependent and signal-independent components~\cite{image_sensors_book}. Most approaches propose to model these noises with some parametric model. For example, ELD~\cite{wei2021physics} identifies four key noise elements in RAW sensor data: shot noise, row noise, generalized read noise, and quantization noise, and proposes modeling them statistically. Based on it, LED~\cite{jin2023lighting} presents a sensor-agnostic pre-training and finetuning framework. PMN~\cite{feng2023learnability} highlights the effects of fixed-pattern noise and black-level errors in network training. Monakhova~\etal~\cite{monakhova2022dancing} propose a temporal noise model with periodic and time-variant row noise for starlight video denoising.  
Cao~\etal~\cite{cao2023physics} develop a comprehensive model connecting noise components to camera ISO via a normalizing flow framework, addressing parameter estimation. Zhang~\etal~\cite{zhang2023towards} introduce a generative network for signal-independent noise synthesis. PNNP~\cite{feng2023physics} proposes to decouple signal-independent noise to independent components and profile the i.i.d one with a deep proxy network. On the other hand, there are some explorations toward noise synthesis without parametric modeling. For example, Zhang~\etal~\cite{zhang2021rethinking} propose using dark frames directly sampled from the sensors as signal-independent noise. Mosleh~\etal~\cite{Mosleh2024nonparam} leverage histograms to achieve non-parametric noise modeling.  

Yet, since most of these approaches rely on precise system gain for signal-dependent shot noise synthesis, a laborious system gain calibration process is always required. Meanwhile, profiling signal-independent noise with statistical parametric models entails implicit assumptions about noise components based on empirical analyses that would vary among different sensor models, and network-based profiling increases the complexity significantly. Therefore, in this work, we systematically investigate the necessity of calibrating system gain and establish improved practices for characterizing signal-independent noise.

\begin{figure*}[t]
    \centering
    \includegraphics[width=\linewidth]{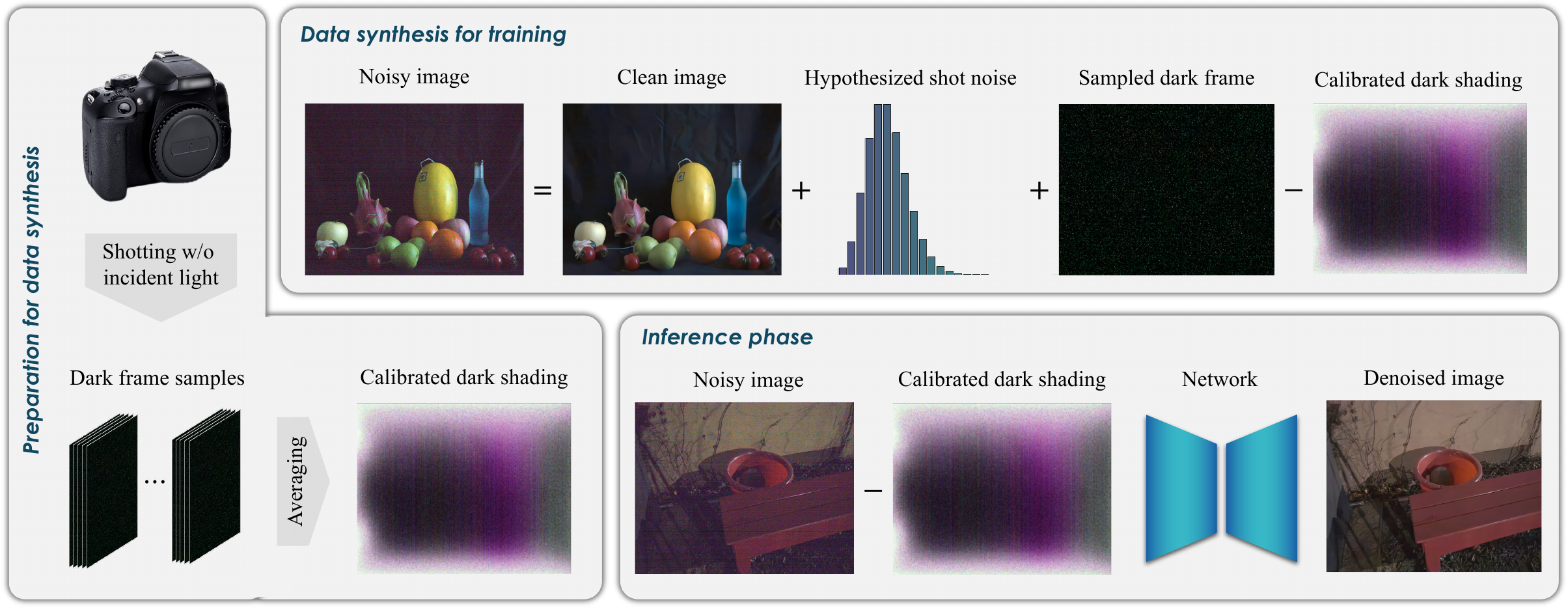}
    \caption{An overview of our denoising pipeline. During the preparation stage, we collect multiple dark frames for each analog gain and compute the corresponding dark shading. To synthesize a pair of training images, we hypothesize a system gain $K$ to generate the Poisson shot noise map and add it, together with a sampled dark shading-corrected dark frame, to a clean image. In inference, given a noisy image, we subtract the dark shading from it to input to the network.}
    \label{fig:overview}
\end{figure*}

\section{Toward practically simple noise synthesis}\label{sec:method}
Given a clean RAW image, our goal lies in synthesizing a photorealistic low-light noisy counterpart that aligns with a specific sensor model. Here, we show how to achieve this while minimizing the commonly required preparation efforts. An overview of our proposal is illustrated in \fref{fig:overview}.

\subsection{Image formation model}
We follow the established noisy image formation model~\cite{wei2021physics,wang2020practical} in this work, where the conversion from incident photons to a digital pixel value can be expressed as:
\begin{equation}
    D = K_d\left(K_a\left(X + N_p + N_1 \right) + N_2\right),  
    \label{eq:img_formation_model}
\end{equation}
where the analog-to-digital conversion (ADC) step is omitted for simplicity, $K_d$ is a digital gain applied for brightness adjustment, $K_a$ is the system gain applied to the analog signal, $X$ is the real number of photons proportional to the scene irradiance, $N_p$ is the signal-dependent noise, and $N_1$ and $N_2$ are two types of signal-independent noise that occurs before and after the ADC, respectively. Without loss of generality, we here assume $K_d=1$ for simplicity, leading to a re-written image formation model of
\begin{equation}
    D = \underbrace{K_a\left(X + N_p\right)}_{\mathrm{signal-dependent}} + \underbrace{K_aN_1 + N_2}_{\mathrm{signal-independent}}.
    \label{eq:img_formation_model_rewrite}
\end{equation}

Among the three types of noises in \eref{eq:img_formation_model_rewrite}, $N_p$ is dominated by the photon-shot noise triggered by the quantum property and inherent uncertainty of light. $N_1$ refers to noise that arises before the functioning of the real-out circuits, such as the dark-current noise and thermal noise, and $N_2$ corresponds to noise that happens at the end of the image formatting process, such as banding noise triggered by non-uniformity of read-out circuits, and quantization noise resulted from bit-depth adjustment.

\subsection{Hypothesizing signal-dependent noise}\label{subsec:hypothesis_k}
The signal-dependent noise is typically simplified as photon-shot noise given its predominance, which imposes a Poisson distribution $\mathcal{P\left(\cdot\right)}$ in the form of
\begin{equation}
    \left(X + N_p\right) \sim \mathcal{P}\left(X\right).
    \label{eq:photon_shot_noise}
\end{equation}
Furthermore, we can augment \eref{eq:photon_shot_noise} into the form of~\footnote{We follow existing works~\cite{wang2020practical, zhang2021rethinking, jin2023lighting} and use $K\mathcal{P}\left(X\right)$ to notate the distribution of scaled Poisson variables, although it is not strict in math.}
\begin{align}
    & \left(X + N_p\right) \sim \mathcal{P}\left(I / K_a\right) \\
   \Rightarrow \quad & K_a\left(X + N_p\right) \sim K_a\mathcal{P}\left(I / K_a\right),
   \label{eq:photon_shot_noise_img}
\end{align}
where $I = K_a X$ is the underlying clean image without any noise. Chiefly, \Eref{eq:photon_shot_noise_img} illustrates the process of generating photon-shot noisy images from their clean counterparts: \ie, sampling from the Poisson distribution \hbox{$\mathcal{P}\left(I / K_a\right)$} and scaling it by $K_a$.

Although \eref{eq:photon_shot_noise_img} appears straightforward, its practical implementation demands considerable effort. Specifically, calibrating the system gain $K$ is often a laborious process involving collecting flat-field images or standard noise test charts (\eg, ISO-15739~\cite{iso9001}) under carefully regulated lighting conditions, as illustrated in \fref{fig:teaser}. Therefore, we seek to bypass the calibration process to ease practical deployment. Our proposal is built upon two observations:


\paragraph{Observation 1: The possible system gain of a certain setup varies in a small range} This observation arises from the fact that the system gain $K$ is composed of the quantum efficiency (QE) and the analog gain (AG)~\cite{european2010standard}, expressed as:
\begin{equation}
    \text{System gain $K$} = \text{QE} \times \text{AG}.
\end{equation}
Of these, AG is a user-designated (thus known) value functioning from pixel-level active transistors to the read-out circuits, and QE represents the overall full-spectral photon-to-electron conversion efficiency, which typically ranges from $30\%$ to $70\%$ under the current advancements in material science (see the supplementary material for more details). Consequently, for a given AG, the maximum and minimum possible values of the corresponding $K$ would only differ by approximately a factor of two.

\paragraph{Observation 2: $K$ only affects noise severity without introducing any intensity bias} This observation stems from a further derivation of \eref{eq:photon_shot_noise_img}. Specifically, for a noisy image sample \hbox{$K\left(X+N_p\right)$} with \hbox{$\left(X+N_p\right) \sim \mathcal{P}\left(I/K\right)$}, its expectation $\mathbb{E}$ and variance $\mathbb{V}$ can be calculated as:
\begin{align}
\begin{split}
    \mathbb{E}\left(K\left(X+N_p\right)\right) &= K \cdot I / K = I, \\
    \mathbb{V}\left(K\left(X+N_p\right)\right) &= K^2  \cdot I / K = K I. 
\label{eq:mean_var_of_shot_noise}
\end{split}
\end{align}
Chiefly, \Eref{eq:mean_var_of_shot_noise} suggests that noisy images generated with distinct $K$ values will only differ in noise severity (\ie, the variance is $K$-scaled), and they will not exhibit any relative biases in pixel intensities as the mean value is independent of $K$.

\vspace{1em}
\noindent Based on the two observations mentioned above, we propose a hypothesizing-based method for shot noise synthesis. Specifically, for a given AG, we start by hypothesizing a QE value (\eg, \hbox{$\left(30\%+70\%\right)/2 = 50\%$)}, multiplying it with the AG to obtain the system gain $K$, and applying \eref{eq:photon_shot_noise_img} to generate the shot-noisy image. Since the hypothesized $K$ is physically constrained to remain close to the underlying ground truth, using it can be intuitively viewed as a form of data augmentation by perturbing AG. This is particularly well-suited for the practical scenario where a single network is trained to handle a broad range of AG values associated with the sensor. Moreover, since different $K$ values only result in noise severity variation, a denoising network, even trained with imprecise $K$, is consistently free from learning any offsets to regress toward the statistical expectation (\ie, the clean image $I$), thereby preventing undesired color biases during inference.

\subsection{Sampling signal-independent noise}\label{subsec:signal_independent_noise}
The dominant approaches for signal-independent noise synthesis involve profiling the noise with statistical models~\cite{feng2023physics,wei2021physics, wang2020practical, cao2023physics, jin2023lighting}. However, we consider accurate profiling rarely possible given the complex noise distribution, and any attempts~\cite{abdelhamed2019noise,wei2021physics,wang2020practical} to refine the modeling process would inevitably add further implementation complexity and concerns in bias-variance trade-off. For smooth text flow, we omit further details here and present a comprehensive justification of statistical noise profiling later in \Sref{subsec:analyze_signal_ind_noise}.

To address the aforementioned issues, in this work, we opt to sample signal-independent noise directly from the sensors. Specifically, following SFRN~\cite{zhang2021rethinking}, we capture multiple dark frames with the target sensor and employ them as direct samples of the overall signal-independent noise, instantiating \hbox{$k_aN_1 + N_2$} in \eref{eq:img_formation_model_rewrite}. In practice, this capturing process can be completed in a few hours in a fully automatic manner. 

To ease the training process, we further bypass the effects of temporally consistent noise via dark shading correction, as popularly conducted in professional photography~\cite{buchheim2007sky, covington1999astrophotography} and RAW image denoising works~\cite{abdelhamed2019noise, feng2023learnability, cao2023physics}. This process involves calibrating system gain-specific dark shadings by averaging multiple dark frames, and subtracting them from the noisy images before inputting to the network. In our case, such a calibration process can be straightforwardly achieved by reusing dark frames collected for noise synthesis, and thus requires no additional effort for preparation.

A key difference between our work and the existing ones~\cite{feng2023learnability, zhang2021rethinking, feng2023physics} lies in the usage of high-bit-depth noise recovery (HBNR), 
a core process designed to mitigate the effects of quantization noise by expanding the bit-depth of dark frames. 
This technique operates by first profiling the signal-independent noise with several simple distributions (Gaussian, Uniform, \etc), selecting the best fit, and then perturbing the original low-bit dark frames based on the corresponding percent-point function. As a result, HBNR adds significant further complexity to noise synthesis by re-necessitating statistical signal-independent noise profiling and sensor-specific justification. Moreover, based on two further considerations, we have chosen to omit HBNR from our pipeline: (1) We are concerned that HBNR may introduce additional errors due to discrepancies between the underlying true noise distribution and the fitted simple ones. (2) We consider the low-bit property of dark frames, rather than being a disadvantage, actually serves as a realistic depiction of the quantization noise. This is because all noisy images provided during inference would be in the low-bit format with the quantization noise implicitly encoded. An empirical study \wrt HBNR in the context of our noise synthesis pipeline can be found in \Sref{subsec:analyze_signal_ind_noise}.

\begin{table*}[t]
\centering
\begin{adjustbox}{max width=\linewidth}
\addtolength{\tabcolsep}{-0.4em}
\begin{NiceTabular}{cccccc}
    & Noisy input & ELD~\cite{wei2021physics} & SFRN~\cite{zhang2021rethinking} & Ours & Ground truth \\
    \raisebox{0.6cm}{\rotatebox{90}{SID dataset}} &
    \includegraphics[width=0.23\textwidth, height=3cm]{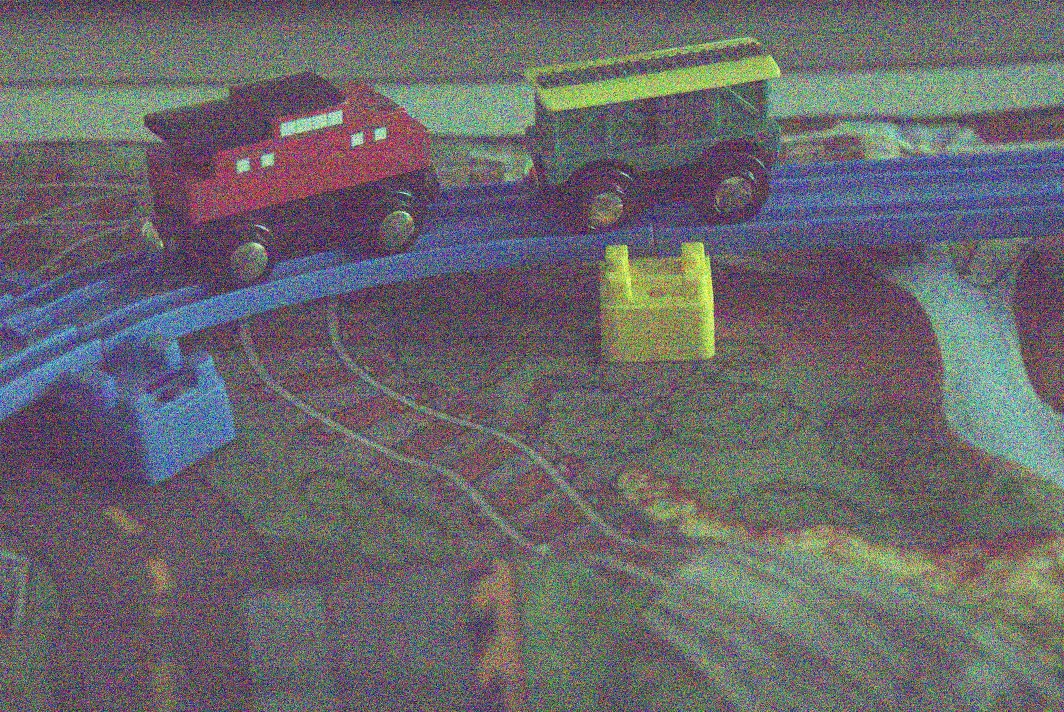} & 
    \includegraphics[width=0.23\textwidth, height=3cm]{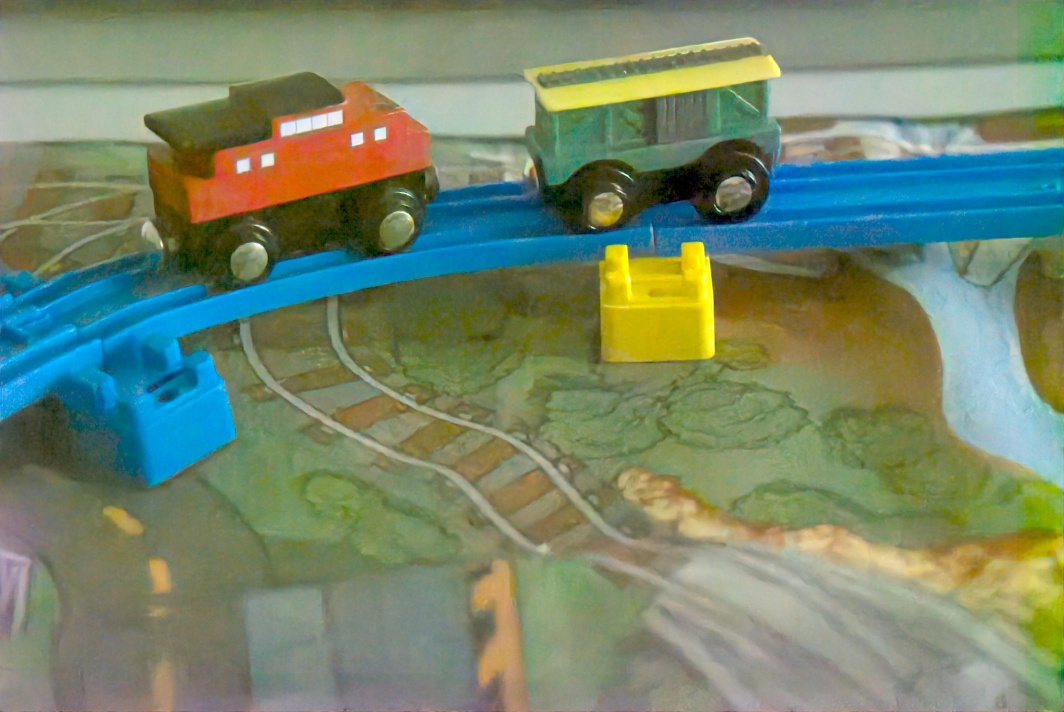} & 
    \includegraphics[width=0.23\textwidth, height=3cm]{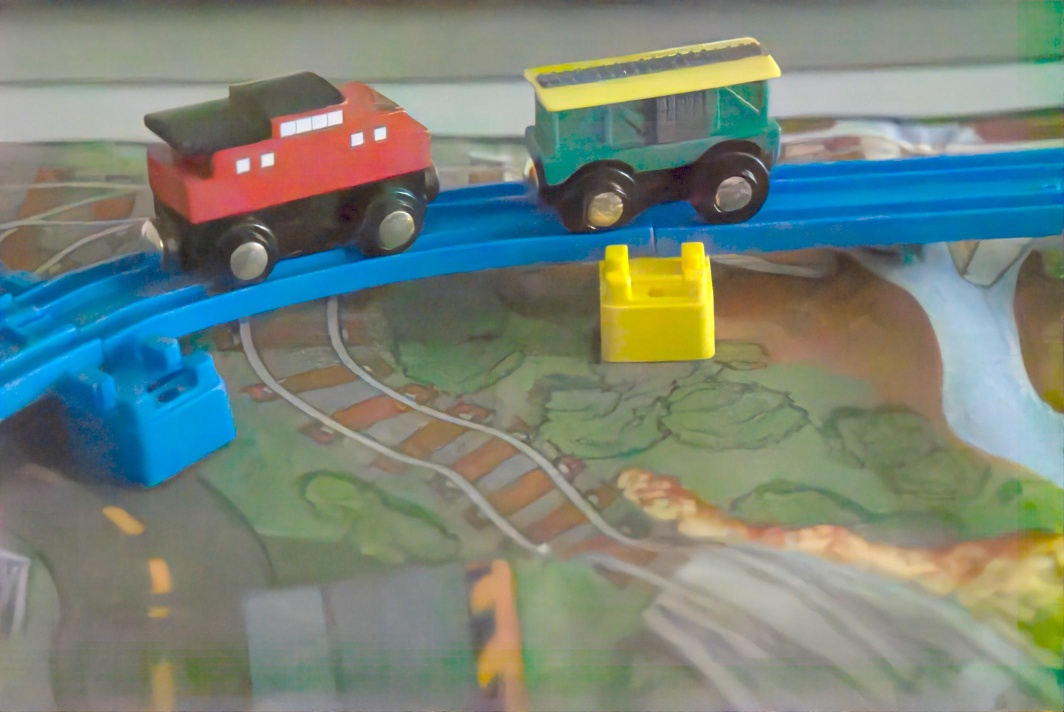} & 
    \includegraphics[width=0.23\textwidth, height=3cm]{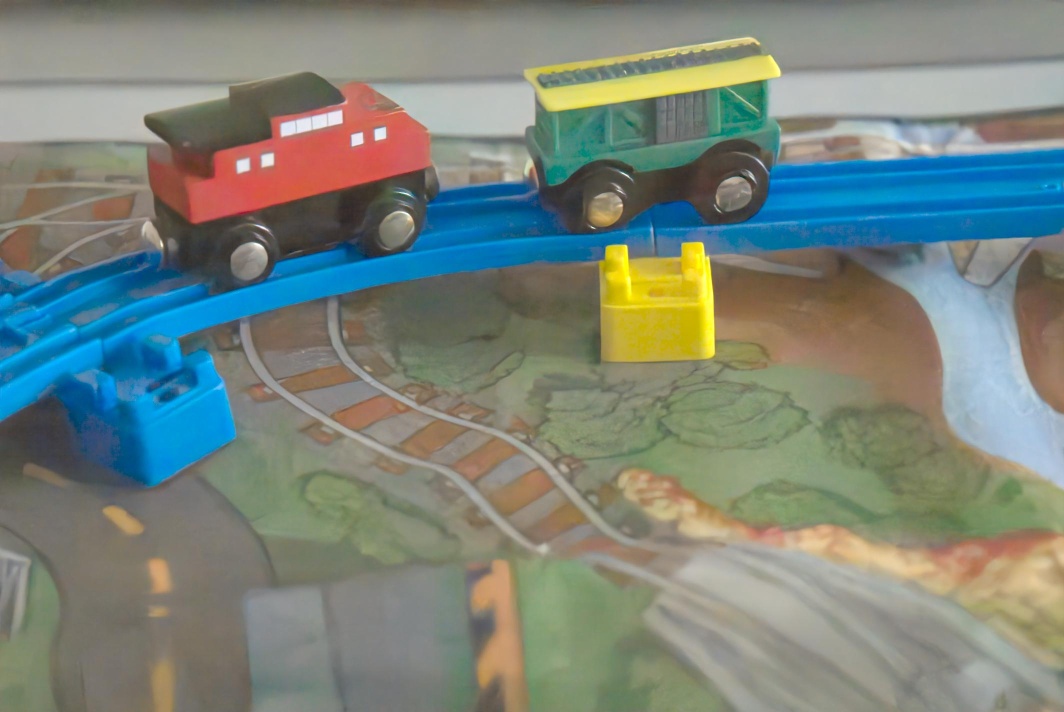} &  
    \includegraphics[width=0.23\textwidth, height=3cm]{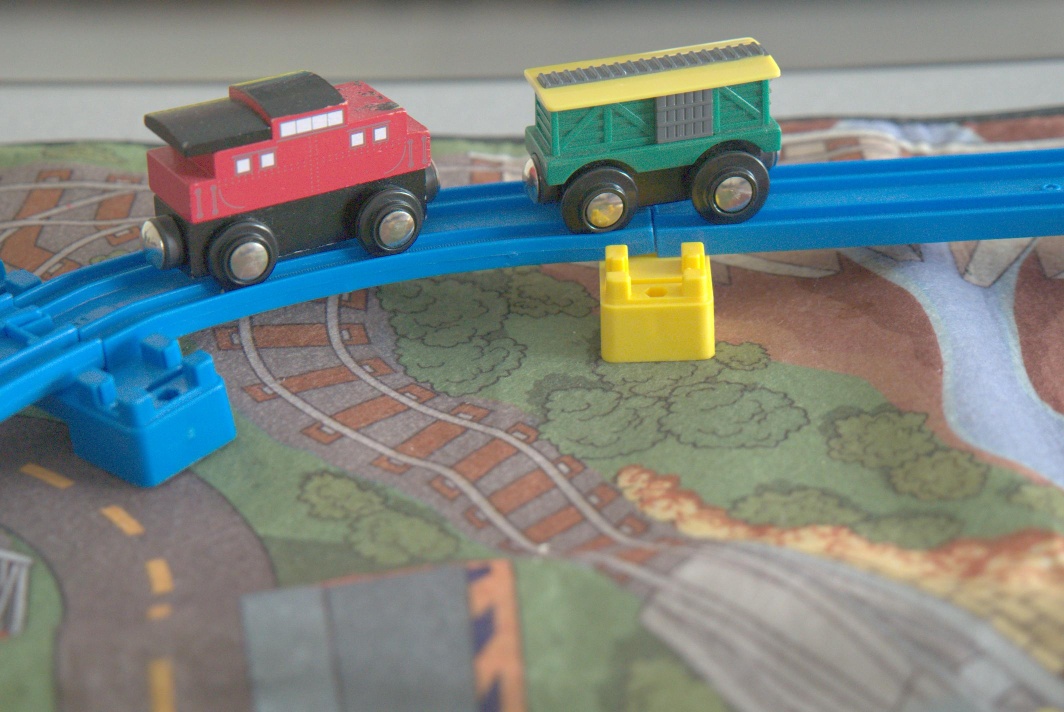} \\

    \raisebox{0.6cm}{\rotatebox{90}{ELD dataset}} &
    \includegraphics[width=0.23\textwidth, height=3cm]{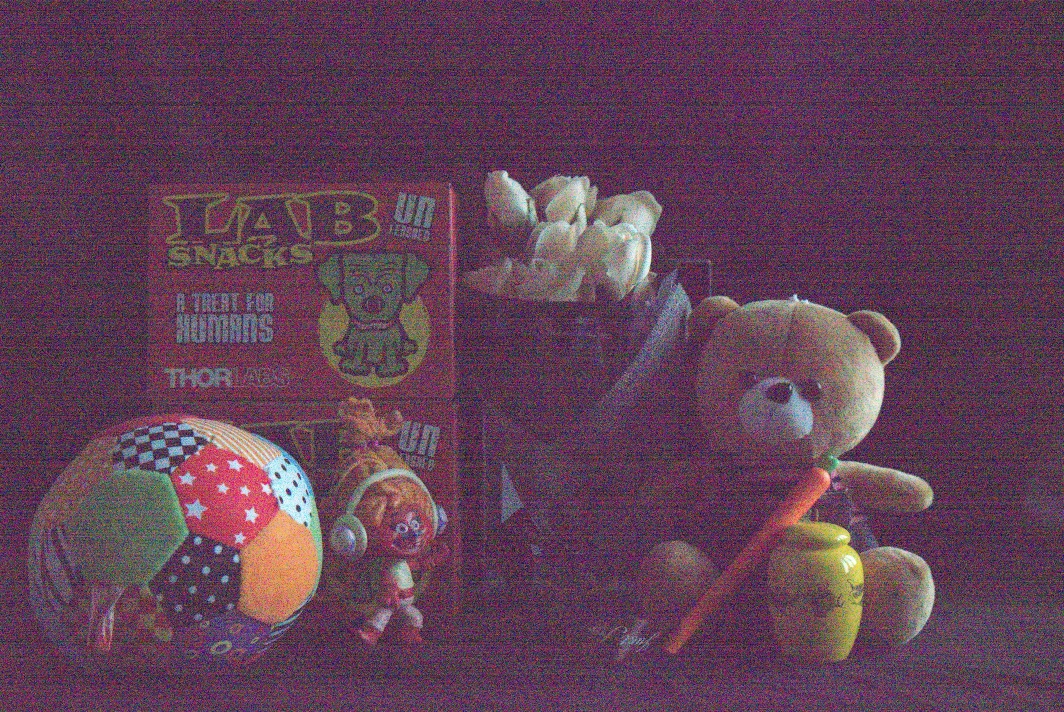} & 
    \includegraphics[width=0.23\textwidth, height=3cm]{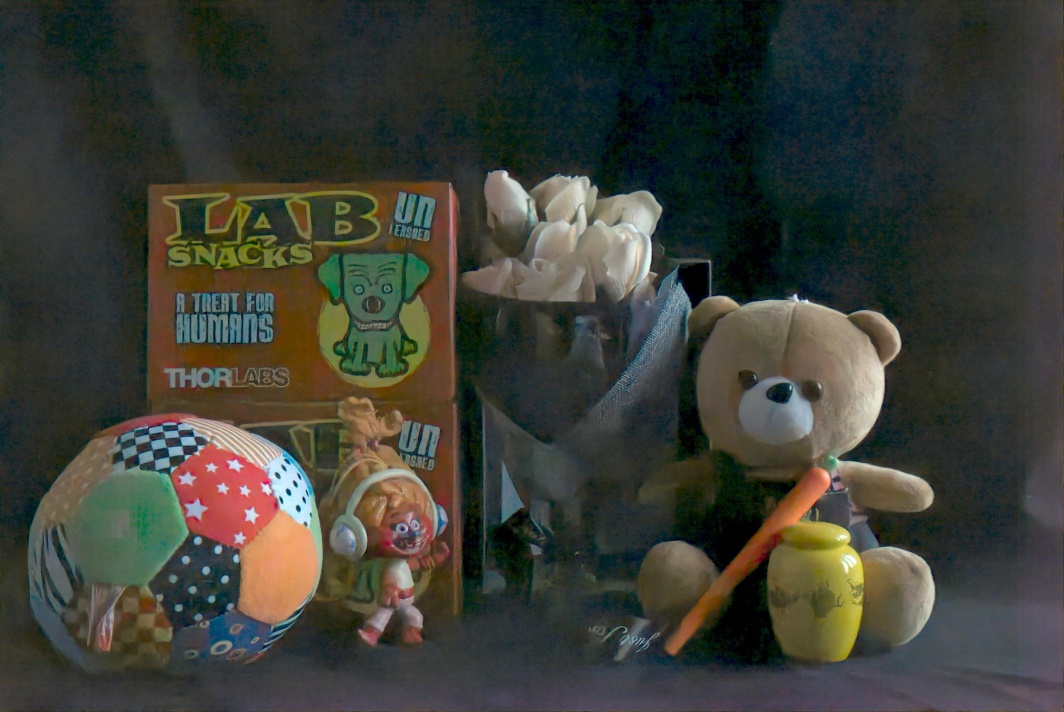} & 
    \includegraphics[width=0.23\textwidth, height=3cm]{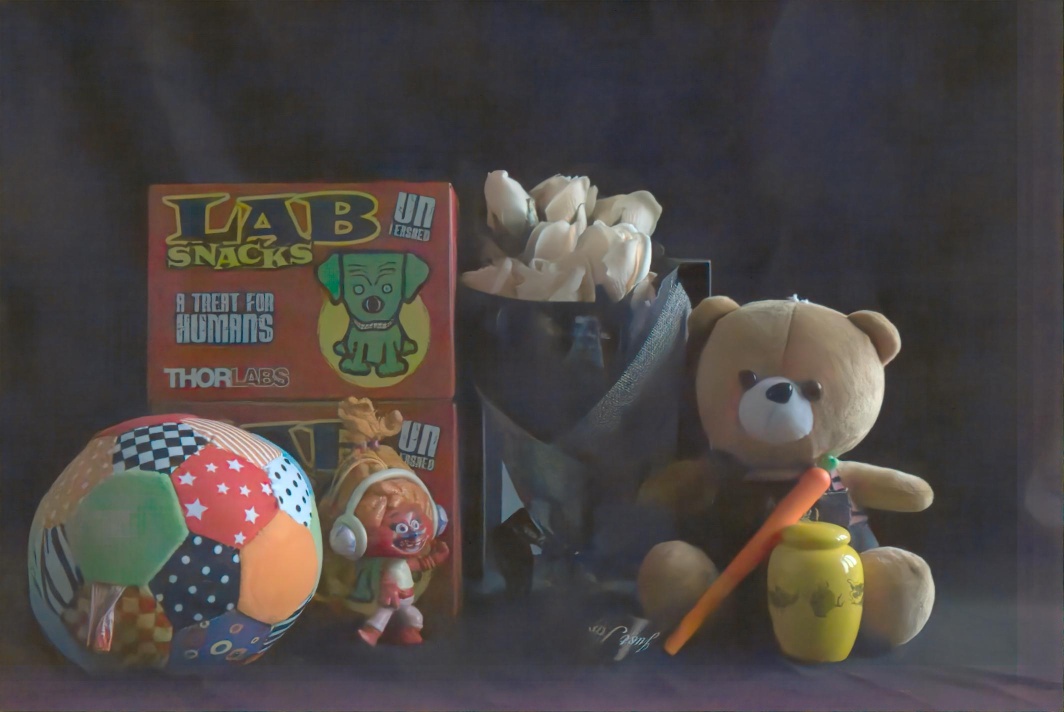} & 
    \includegraphics[width=0.23\textwidth, height=3cm]{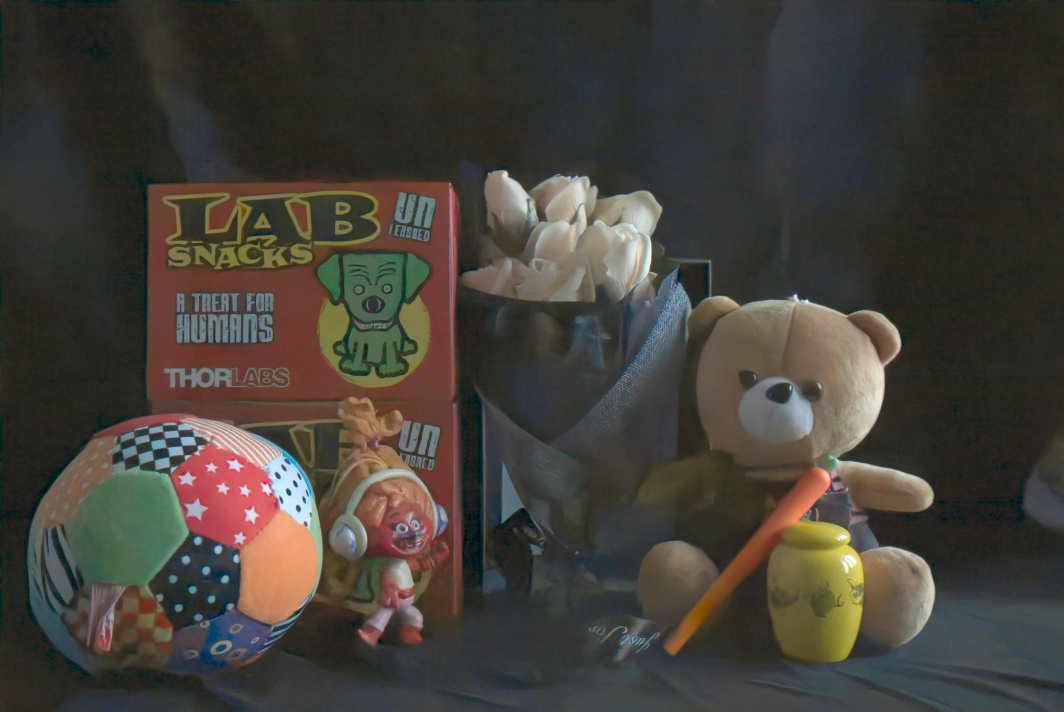} &  
    \includegraphics[width=0.23\textwidth, height=3cm]{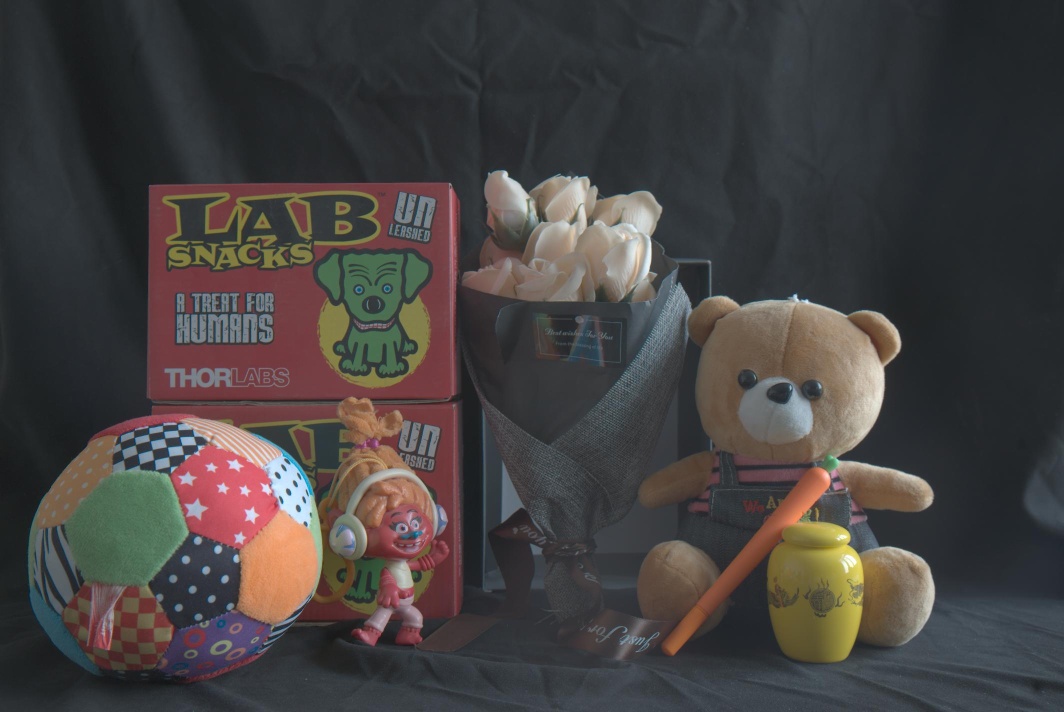} \\

    \raisebox{0.6cm}{\rotatebox{90}{LRID dataset}} &
    \includegraphics[width=0.23\textwidth, height=3cm]{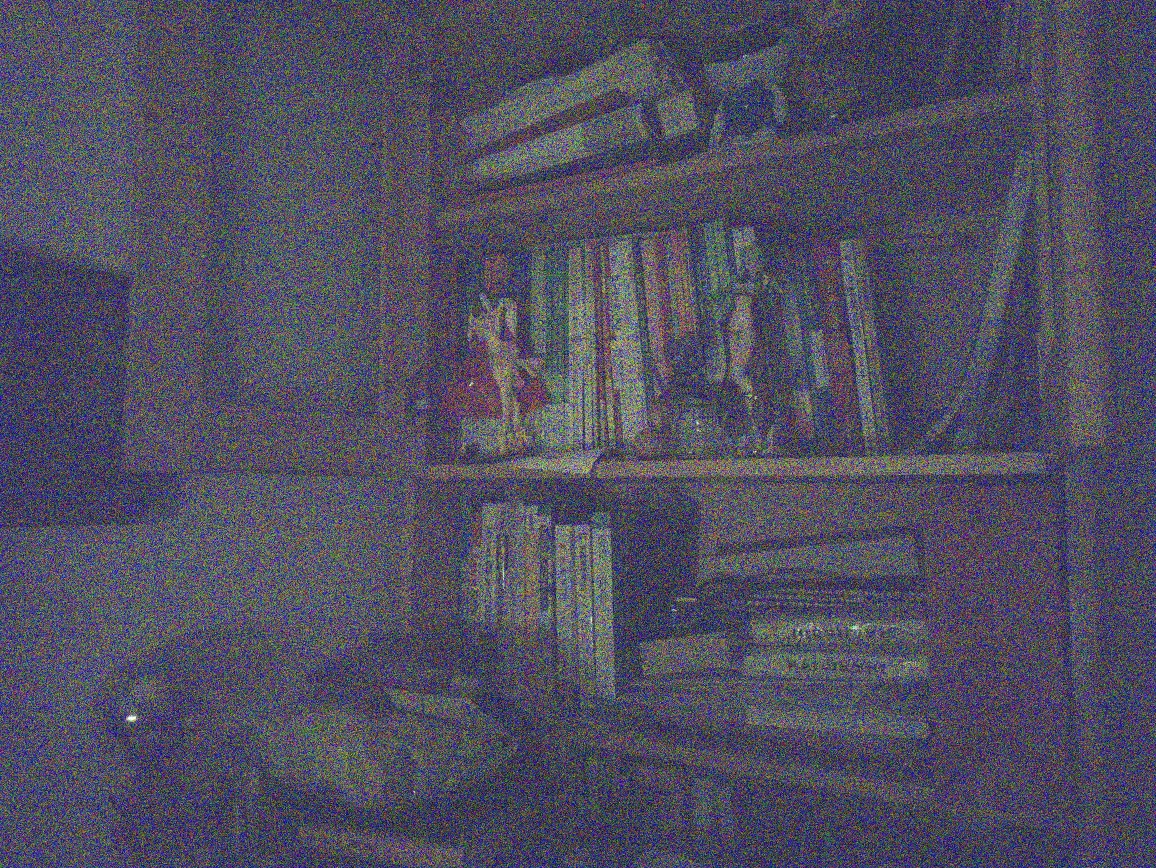} & 
    \includegraphics[width=0.23\textwidth, height=3cm]{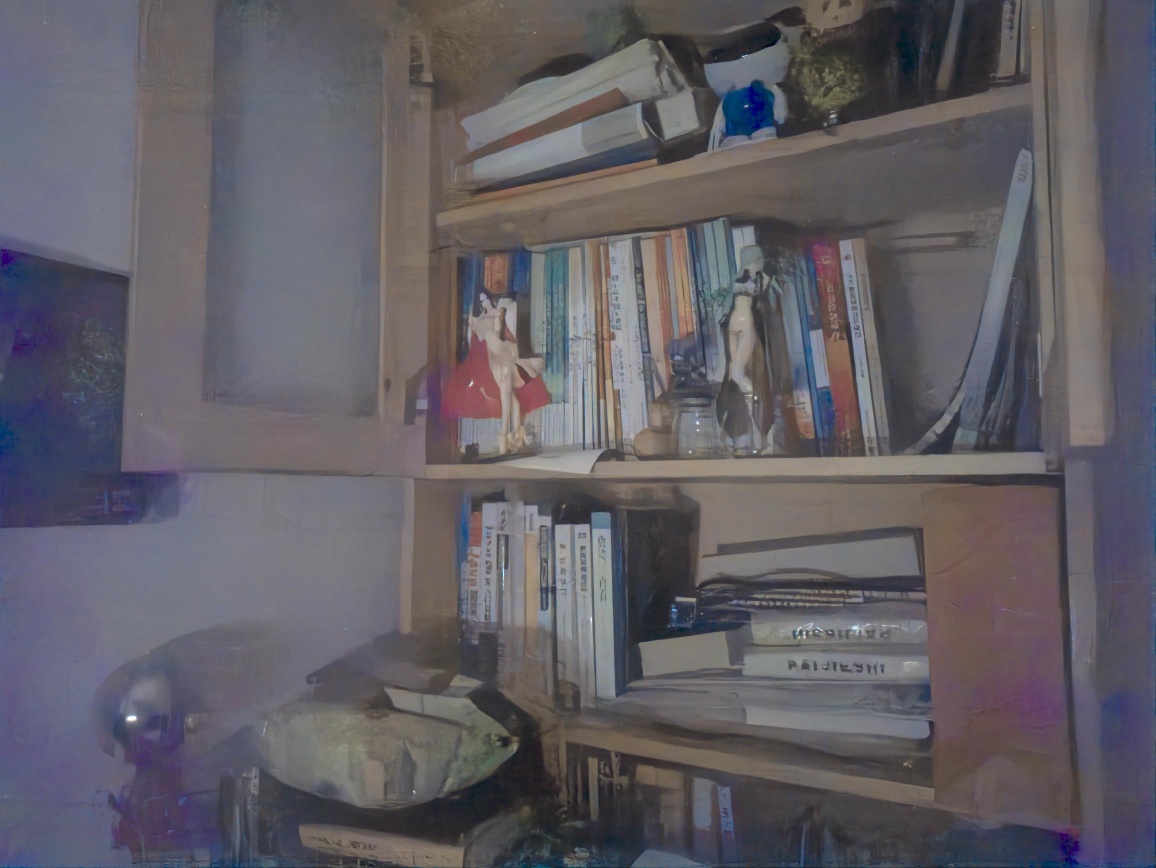} & 
    \includegraphics[width=0.23\textwidth, height=3cm]{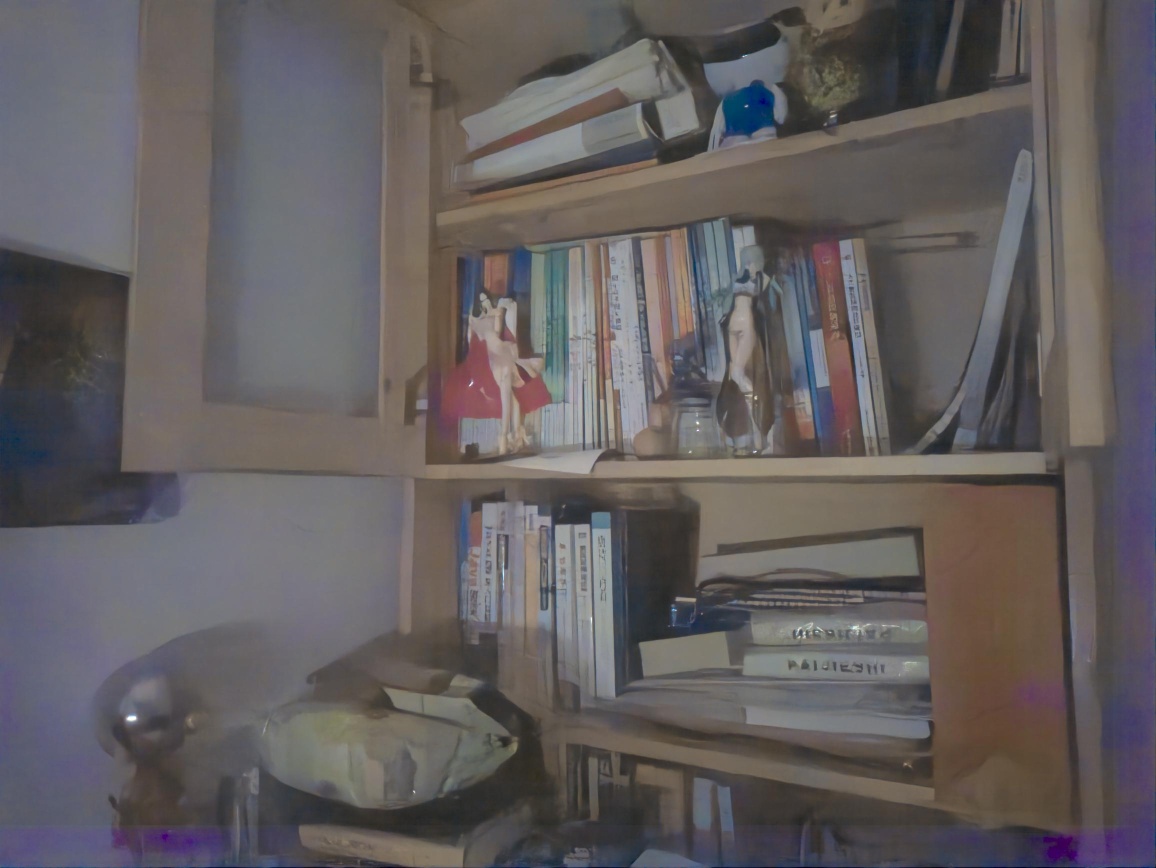} & 
    \includegraphics[width=0.23\textwidth, height=3cm]{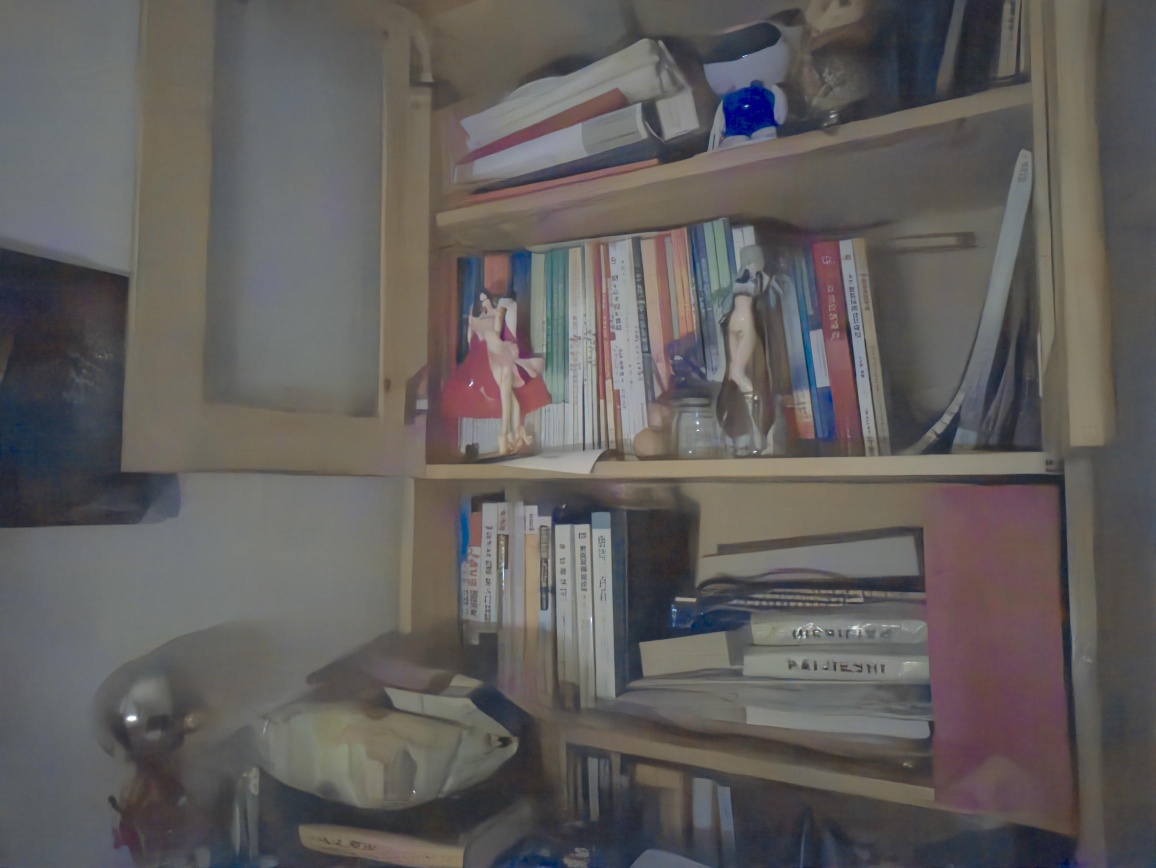} &  
    \includegraphics[width=0.23\textwidth, height=3cm]{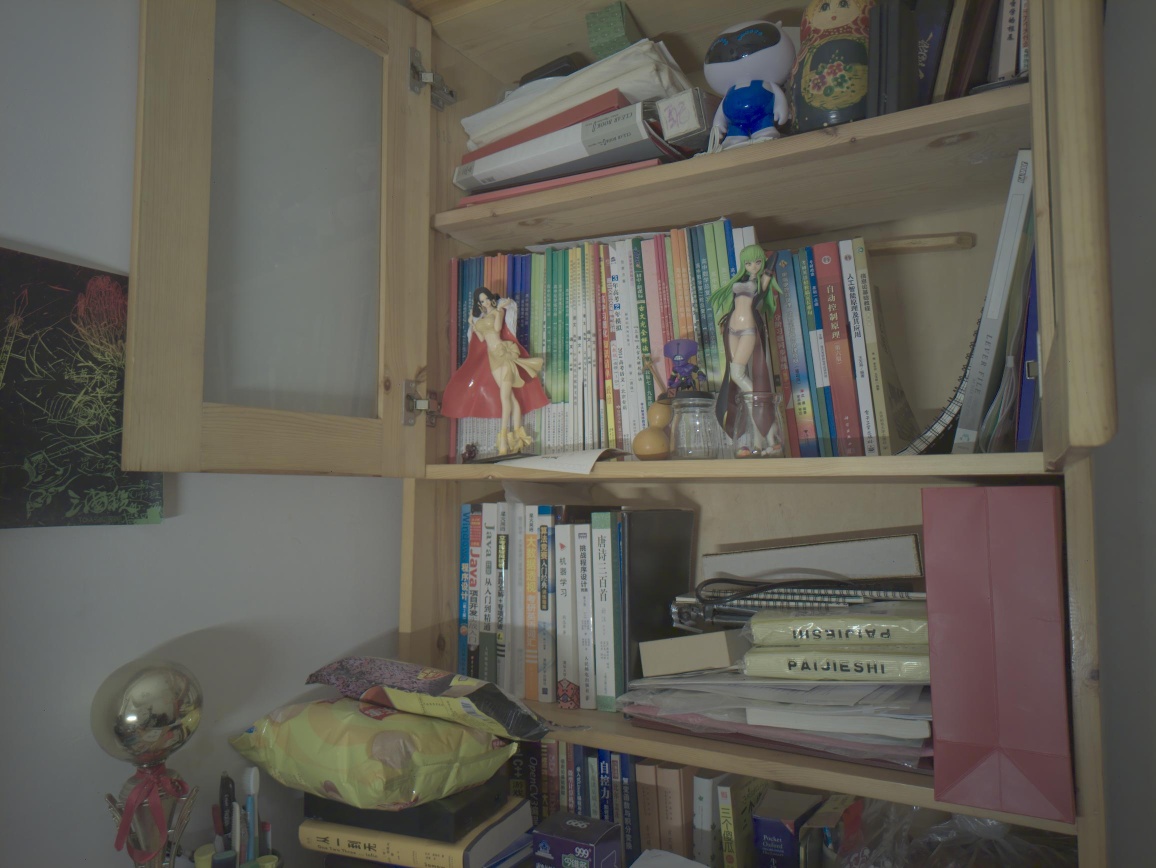} \\

\end{NiceTabular}
\end{adjustbox}
\captionof{figure}{Qualitative comparison of distinct noise synthesis approaches. All images are processed to sRGB via a simple image signal processing pipeline for better visualization. ELD and SFRN exhibit noticeable color bias along image boundaries, whereas ours does not.}
\label{fig:denoise_results}
\end{table*}

\section{Experiments}\label{sec:experiment}
We extensively compare our proposal with state-of-the-art methods and present comprehensive ablation studies.

\subsection{Experimental setup}\label{subsec:expt_setup}
\paragraph{Datasets} We employ three popular datasets for low-light raw image denoising: SID~\cite{chen2018learning}, ELD~\cite{wei2021physics}, and LRID~\cite{feng2023learnability}. Among them, SID and ELD are captured with two different Sony-A7S2 full-frame DSLR cameras, and LRID is collected by a Redmi K30 smartphone with a $1/2.55$ inch IMX686 sensor. Among them, SID and ELD provide test data with multiple ISO values ranging from $100$ to $25600$, and LRID specifies ISO $6400$ for the test data. For noise synthesis, since neither SID nor ELD has released dark frames, we leverage the LLD~\cite{cao2023physics} calibration data (approx. $400$ dark frames per ISO, $24$ ISO in total) to aid experiments on SID and LED. As for LRID, we employ the $360$ ($180$ normal and $180$ hot) dark frames provided as part of the dataset.

\paragraph{Comparison methods} We involve most of the SoTA methods on low-light RAW image denoising into comparison, including supervised methods that solely rely on real-world noisy-clean image pairs, self-supervised methods that synthesize noise based on clean images, and the hybrid ones that use synthetic data for pre-training and real-world image pairs for finetuning. Specifically, we compare to
\begin{itemize}
    \item \textbf{Supervised}: Vanilla paired data and PMN~\cite{feng2023learnability}.
    \item \textbf{Self-supervised}: Poisson-Gaussian (P-G), SFRN~\cite{zhang2021rethinking}, ELD~\cite{wei2021physics}, NoiseFlow~\cite{abdelhamed2019noise}, StarLight~\cite{monakhova2022dancing}, and the PNNP~\cite{feng2023physics} method developed concurrently with us. 
    \item \textbf{Hybrid}: LED~\cite{jin2023lighting} and LLD~\cite{cao2023physics}.
\end{itemize}
All these approaches have employed the same U-Net specified in~\cite{chen2018learning} as the denoising backbone, and we keep in line with them for fair comparisons.

\paragraph{Training and test specifications} For training, we align with existing works~\cite{feng2023learnability, cao2023physics, jin2023lighting} to randomly crop the images and dark frames to $512\times512$ patches, and minimize an L1-loss between the predictions and ground truths with an Adam optimizer~\cite{kingma2014adam} for $1000$ epochs with $400$ epochs for finetuning. Random vertical and horizontal flipping are employed for data augmentation. Regarding the selection of the overall system gain $K$, since we do not have any information about how the two devices used for data collection encapsulate analog gains to ISO values, we empirically investigate their characteristics and find that both cameras have their base ISOs (\ie, $\text{AG}=1$) around $400$. To this end, we employ an empirical approximation equation for the system gain $K$: \hbox{$K = \mathrm{ISO} / 100 * 0.1$}, representing a hypothesized QE of $40\%$. For testing, we report the PSNR and SSIM scores in line with existing works. Throughout the training and testing phases, we consistently use the dark shadings provided by PMN~\cite{feng2023learnability} for a fair comparison.

\subsection{Denoising performance}\label{subsec:denoise_performance}

\paragraph{SID and ELD datasets}
Some qualitative results are illustrated in the first two rows of \fref{fig:denoise_results} and the quantitative ones are reported in \Tref{tab:sid_eld_res}. Our proposed noise synthesis pipeline achieves comparable results to the concurrent approach PNNP~\cite{feng2023physics} while avoiding its tedious procedure of training an auxiliary noise synthesis network and system gain calibration, and outperforms the other methods by a large margin. Moreover, our self-supervised method can even outperform the state-of-the-art supervised method PMN~\cite{feng2023learnability} due to the imperfect pixel alignments of the SID noisy-clean training pairs. The above two phenomena emphasize the practical benefits of our proposal in achieving easy-to-land and high-accuracy RAW image denoising.

\begin{table*}[t]
\centering
\caption{Denoising results (PSNR/SSIM) on the SID~\cite{chen2018learning} and ELD~\cite{wei2021physics} Sony-A7S2 datasets. \red{Red} and \blue{blue} refer to the best and second-best results, respectively.}
\begin{adjustbox}{max width=\linewidth}
\begin{NiceTabular}{c|c|cccc|ccc}
\toprule
    \multirow{2}{*}{\textbf{\makecell{Training setup}}} & 
    \multirow{2}{*}{\textbf{Method}} &
    \multicolumn{4}{c|}{\textbf{SID}} &
    \multicolumn{3}{c}{\textbf{ELD}} \\
\cmidrule{3-6}
\cmidrule{7-9}
    & & 
    $\V{\times100}$ & 
    $\V{\times250}$ & 
    $\V{\times300}$ & 
    \textbf{Avg.} & 
    $\V{\times100}$ & 
    $\V{\times200}$ & 
    \textbf{Avg.} \\    
\midrule
    \multirow{2}*{\makecell{Supervised}} 
    & \textbf{Paired} & 42.06 / 0.9548 & 39.60 / 0.9380 & 36.85 / 0.9227 & 39.32 / 0.9374 & 44.47 / 0.9676 & 41.97 / 0.9282 & 43.22 / 0.9479 \\

    & \textbf{PMN~\cite{feng2023learnability}} & 43.47 / 0.9606 & 41.04 / 0.9471 & 37.87 / 0.9344 & 40.59 / 0.9465 & 46.99 / 0.9840 & 44.85 / 0.9686 & 45.92 / 0.9763 \\
\cmidrule{1-9}
    \multirow{2}*{\makecell{Hybrid}} 
    & \textbf{LED~\cite{jin2023lighting}} & 41.98 / 0.9539 & 39.34 / 0.9317 & 36.67 / 0.9147 & 39.19 / 0.9321 & 45.36 / 0.9779 & 42.97 / 0.9577 & 44.17 / 0.9678 \\
    & \textbf{LLD~\cite{cao2023physics}} & 43.36 / 0.9610 & 41.02 / 0.9480 & 37.80 / 0.9350 & 40.52 / 0.9471 & 46.74 / 0.9860 & 44.95 / 0.9770 & 45.85 / 0.9815 \\
\cmidrule{1-9} 
    \multirow{7}*{\makecell{Self-\\supervised}} 
    & \textbf{P-G} & 39.44 / 0.8995 & 34.32 / 0.7681 & 30.66 / 0.6569 & 34.52 / 0.7666 & 42.05 / 0.8721 & 38.18 / 0.7827 & 40.12 / 0.8274 \\

    & \textbf{ELD~\cite{wei2021physics}} & 41.95 / 0.9530 & 39.44 / 0.9307 & 36.36 / 0.9114 & 39.05 / 0.9303 & 45.45 / 0.9754 & 43.43 / 0.9544 & 44.44 / 0.9649 \\

    & \textbf{SFRN~\cite{zhang2021rethinking}} & 42.61 / 0.9580 & 40.73 / 0.9454 & 37.64 / 0.9309 & 40.14 / 0.9438 & 46.45 / 0.9843 & 44.58 / 0.9738 & 45.51 / 0.9790 \\

    & \textbf{NoiseFlow~\cite{abdelhamed2019noise}} & 41.08 / 0.9394 & 37.45 / 0.8864 & 33.53 / 0.8132 & 37.09 / 0.8750 & 43.21 / 0.9210 & 40.60 / 0.8638 & 41.90 / 0.8924 \\
    
    & \textbf{StarLight~\cite{monakhova2022dancing}} & 40.47 / 0.9261 & 36.26 / 0.8575 & 33.00 / 0.7802  & 36.33 / 0.8494 & 43.80 / 0.9358 & 40.86 / 0.8837 & 42.33 / 0.9098 \\

    & \textbf{PNNP~\cite{feng2023physics}} & 
    \textbf{\blue{43.63 / 0.9614}} & 
    \textbf{\red{41.49} / \red{0.9498}} & 
    \textbf{\blue{38.01 / 0.9353}} & 
    \textbf{\blue{40.83} / \red{0.9479}} & 
    \textbf{\blue{47.31} / \red{0.9877}} & 
    \textbf{\blue{45.47} / \blue{0.9791}} &
    \textbf{\blue{46.39} / \red{0.9834}} \\ 

    & \textbf{Ours} & 
    \textbf{\red{43.69 / 0.9618}} & 
    \textbf{\blue{41.43} / \blue{0.9486}} & 
    \textbf{\red{38.06 / 0.9356}} & 
    \textbf{\red{40.85} / \blue{0.9478}} & 
    \textbf{\red{47.34} / \blue{0.9874}} & 
    \textbf{\red{45.51} / \red{0.9794}} & 
    \textbf{\red{46.43} / \red{0.9834}} \\ 

\bottomrule
\end{NiceTabular}
\end{adjustbox}
\label{tab:sid_eld_res}
\end{table*}

\begin{table*}[t]
\centering
\caption{Denoising results (PSNR/SSIM) on the LRID dataset~\cite{feng2023learnability}. \red{Red} and \blue{blue} refer to the best and second-best results, respectively.}
\begin{adjustbox}{max width=\linewidth}
\begin{NiceTabular}{c|c|cccccc|cccc|c}
\toprule
    \multirow{2}{*}{\textbf{\makecell{\textbf{Training} \textbf{setup}}}} & 
    \multirow{2}{*}{\textbf{Method}} &
    \multicolumn{6}{c|}{\textbf{Indoor}} &
    \multicolumn{4}{c}{\textbf{Outdoor}} &
    \multirow{2}{*}{\makecell{\textbf{Overall} \textbf{avg.}}} \\
\cmidrule{3-12}
    & & 
    $\V{\times64}$ & 
    $\V{\times128}$ & 
    $\V{\times256}$ & 
    $\V{\times512}$ &
    $\V{\times1024}$ &
    \textbf{Avg.} &  
    $\V{\times64}$ & 
    $\V{\times128}$ & 
    $\V{\times256}$ &
    \textbf{Avg.} \\
\midrule
    \multirow{3}*{\makecell{Supervised}} & 
    \textbf{Paired} & 
    \makecell{48.77 \\ 0.9906} & 
    \makecell{47.00 \\ 0.9860} & 
    \makecell{44.74 \\ 0.9786} & 
    \makecell{42.40 \\ 0.9647} & 
    \makecell{40.07 \\ 0.9437} & 
    \makecell{44.60 \\ 0.9727} & 
    \makecell{45.84 \\ 0.9876} & 
    \makecell{44.50 \\ 0.9821} & 
    \makecell{42.66 \\ 0.9709} & 
    \makecell{44.33 \\ 0.9802} & 
    \makecell{44.52 \\ 0.9748} \\
\cdashedline{2-13}
    & \textbf{PMN~\cite{feng2023learnability}} & 
    \makecell{\textbf{\blue{49.24}} \\ \textbf{\blue{0.9916}}} & 
    \makecell{\textbf{\blue{47.47}} \\ \textbf{\blue{0.9868}}} & 
    \makecell{\textbf{\blue{45.36}} \\ \textbf{\blue{0.9804}}} & 
    \makecell{\textbf{\blue{43.09}} \\ \textbf{\blue{0.9671}}} & 
    \makecell{40.20 \\ 0.9453} & 
    \makecell{\textbf{\blue{45.07}} \\ \textbf{\blue{0.9743}}} & 
    \makecell{\textbf{\red{46.27}} \\ \textbf{\red{0.9884}}} & 
    \makecell{\textbf{\red{44.86}} \\ \textbf{\blue{0.9834}}} & 
    \makecell{\textbf{\red{42.99}} \\ \textbf{\blue{0.9703}}} & 
    \makecell{\textbf{\red{44.71}} \\ \textbf{\blue{0.9807}}} & 
    \makecell{\textbf{\blue{44.97}} \\ \textbf{\blue{0.9761}}} \\
\midrule
    \multirow{11}*{\makecell{Self-\\supervised}} & \textbf{NoiseFlow~\cite{abdelhamed2019noise}} & 
    \makecell{48.16 \\ 0.9901} & 
    \makecell{46.19 \\ 0.9828} & 
    \makecell{43.91 \\ 0.9698} &
    \makecell{41.09 \\ 0.9442} & 
    \makecell{37.76 \\ 0.8906} & 
    \makecell{43.42 \\ 0.9555} & 
    \makecell{45.34 \\ 0.9856} & 
    \makecell{43.82 \\ 0.9757} & 
    \makecell{41.92 \\ 0.9570} & 
    \makecell{43.69 \\ 0.9728} & 
    \makecell{43.72 \\ 0.9596} \\
\cdashedline{2-13}
    & \textbf{P-G} & 
    \makecell{46.14 \\ 0.9872} & 
    \makecell{44.98 \\ 0.9809} & 
    \makecell{43.31 \\ 0.9682} & 
    \makecell{40.80 \\ 0.9429} & 
    \makecell{37.74 \\ 0.8905} & 
    \makecell{42.59 \\ 0.9539} & 
    \makecell{42.16 \\ 0.9796} & 
    \makecell{41.48 \\ 0.9709} & 
    \makecell{40.36 \\ 0.9525} & 
    \makecell{41.33 \\ 0.9677} & 
    \makecell{42.23 \\ 0.9578} \\
\cdashedline{2-13}
    & \textbf{ELD~\cite{wei2021physics}} & 
    \makecell{48.19 \\ 0.9898} & 
    \makecell{46.55 \\ 0.9836} & 
    \makecell{44.39 \\ 0.9730} & 
    \makecell{41.56 \\ 0.9452} & 
    \makecell{37.50 \\ 0.8915} & 
    \makecell{43.64 \\ 0.9566} & 
    \makecell{45.00 \\ 0.9841} & 
    \makecell{43.48 \\ 0.9734} & 
    \makecell{41.31 \\ 0.9450} & 
    \makecell{43.26 \\ 0.9675} & 
    \makecell{43.53 \\ 0.9597} \\
\cdashedline{2-13}
    & \textbf{SFRN~\cite{zhang2021rethinking}} & 
    \makecell{47.94 \\ 0.9899} & 
    \makecell{46.52 \\ 0.9854} & 
    \makecell{44.74 \\ 0.9786} & 
    \makecell{42.46 \\ 0.9652} & 
    \makecell{40.10 \\ 0.9453} & 
    \makecell{44.35 \\ 0.9729} & 
    \makecell{45.05 \\ 0.9850} & 
    \makecell{43.67 \\ 0.9766} & 
    \makecell{41.89 \\ 0.9591} & 
    \makecell{43.54 \\ 0.9736} & 
    \makecell{44.12 \\ 0.9731} \\
\cdashedline{2-13}
    & \textbf{PNNP~\cite{feng2023physics}} & 
    \makecell{48.50 \\ 0.9908} & 
    \makecell{46.94 \\ 0.9863} & 
    \makecell{45.06 \\ 0.9797} & 
    \makecell{42.64 \\ 0.9662} & 
    \makecell{\textbf{\blue{40.30}} \\ \textbf{\blue{0.9460}}} & 
    \makecell{44.69 \\ 0.9738} & 
    \makecell{45.62 \\ 0.9873} & 
    \makecell{44.27 \\ 0.9821} & 
    \makecell{42.63 \\ 0.9724} & 
    \makecell{44.17 \\ 0.9806} & 
    \makecell{44.54 \\ 0.9757} \\
\cdashedline{2-13}
    & \textbf{Ours} & 
    \makecell{\textbf{\red{49.25}} \\ \textbf{\red{0.9918}}} & 
    \makecell{\textbf{\red{47.55}} \\ \textbf{\red{0.9876}}} & 
    \makecell{\textbf{\red{45.53}} \\ \textbf{\red{0.9818}}} & 
    \makecell{\textbf{\red{43.22}} \\ \textbf{\red{0.9695}}} & 
    \makecell{\textbf{\red{40.85}} \\ \textbf{\red{0.9516}}} & 
    \makecell{\textbf{\red{45.28}} \\ \textbf{\red{0.9765}}} &
    \makecell{\textbf{\blue{46.10}} \\ \textbf{\red{0.9884}}} &
    \makecell{\textbf{\blue{44.68}} \\ \textbf{\red{0.9835}}} & 
    \makecell{\textbf{\blue{42.93}} \\ \textbf{\red{0.9719}}} &
    \makecell{\textbf{\blue{44.57}} \\ \textbf{\red{0.9813}}} &
    \makecell{\textbf{\red{45.08}} \\ \textbf{\red{0.9779}}} \\
\bottomrule
\end{NiceTabular}
\end{adjustbox}
\label{tab:lrid_res}
\end{table*}

\paragraph{LRID dataset}
The quantitative results are reported in \Tref{tab:lrid_res}. Our noise synthesis pipeline helps to achieve $0.54\mathrm{dB}$ improvement in PSNR over the state-of-the-art self-supervised method PNNP~\cite{feng2023physics}, and leads to even slightly better performance compared to the fully supervised algorithm PMN~\cite{feng2023learnability}. We attribute this improvement to the fact that the noise patterns would vary among images due to sensor heating, and our employed direct sensor sampling effectively augments the training data to cover such a scenario yet profiling-based methods cannot. A qualitative comparison is presented in the bottom row of \fref{fig:denoise_results}.

\subsection{Ablation studies}\label{subsec:ablation_study}
We here present the robustness of denoising networks \wrt the system gain $K$ in training, and the dark shading accuracy in inference. We also study the data efficiency of our proposed noise synthesis pipeline.

\paragraph{Robustness \wrt the precision of system gain $K$}
As mentioned in \Sref{subsec:hypothesis_k}, the possible range of a system gain $K$ is rather limited due to the physical constraints of the quantum efficiency, and thus $K$ can be safely hypothesized rather than calibrated. To test the network robustness \wrt $K$, we conduct experiments with the following training setups on all three datasets for comprehensiveness:
\begin{itemize}
    \item \textbf{SID and ELD datasets}:
    \begin{itemize}
        \item \textit{Calibrated $K$}: $K$ values calibrated with the flat-field method~\cite{wei2021physics}, resulting a maximum $K\approx 24.48$ corresponding to ISO $25600$.
        \item \textit{Narrow-range hypothesized $K$}: We determine $K$ via $K = \mathrm{ISO} / 100 * 0.1$, resulting a maximum $K$ of $25.6$.
        \item \textit{Broad-range hypothesized $K$}: We determine $K$ via $K = \mathrm{ISO} / 100 * 0.2$, resulting a maximum $K$ of $51.2$.
    \end{itemize}
    \item \textbf{LRID dataset}:
    \begin{itemize}
        \item \textit{Calibrated $K$}: We randomly sample $K$ from $\left(8.7, 8.8\right)$ based on it calibrated value $K\approx8.74$ for ISO $6400$.
        \item \textit{Random $K$}: We randomly sample $K$ from $\left(4, 17\right)$.
    \end{itemize}
\end{itemize}

The results are shown in \Tref{tab:ablation_study_qe}. On all three datasets, the PSNR differences are always smaller than $0.1\mathrm{dB}$ among different training setups. Such results strongly support our claim that denoising networks are robust \wrt the system gain $K$ varying in a wide range, making a hypothesized $K$ based on the quantum efficiency enough and a precise calibration process unnecessary.

\begin{table}[t]
\centering
\caption{Denoising performance \wrt different system gain $K$.}
\label{tab:ablation_study_qe}

\subcaption{SID and ELD datasets}
\begin{adjustbox}{max width=\linewidth}
\begin{NiceTabular}{c|cc}
\toprule
    \textbf{System gain $K$} & \textbf{SID} & \textbf{ELD} \\
\midrule
    Calibrated $K$ & 40.90 / 0.9487 & 46.37 / 0.9832 \\
\midrule
    Narrow-range hypothesized $K$ & 40.85 / 0.9478 & 46.43 / 0.9834 \\
\midrule
    Broad-range hypothesized $K$ & 40.85 / 0.9484 & 46.33 / 0.9826 \\
\bottomrule
\end{NiceTabular}
\end{adjustbox}

\bigskip

\subcaption{LRID dataset}
\begin{adjustbox}{max width=\linewidth}
\begin{NiceTabular}{c|ccc}
\toprule
    \textbf{System gain $K$} & \textbf{Indoor} & \textbf{Outdoor} & \textbf{Avg.} \\
\midrule
    \makecell{Calibrated \\ $K\in\left(8.7, 8.8\right)$} & 
    \makecell{45.20 / 0.9759} & 
    \makecell{44.51 / 0.9809} & 
    \makecell{45.00 / 0.9773} \\
\midrule
    \makecell{Random \\ $K\in\left(4, 17\right)$} & 
    \makecell{45.27 / 0.9764} & 
    \makecell{44.64 / 0.9814} & 
    \makecell{45.09 / 0.9778} \\
\bottomrule
\end{NiceTabular}
\end{adjustbox}
\end{table}

\paragraph{Robustness \wrt inference-phase online dark shading re-calibration}
Generally, dark shadings can be pre-calibrated using the vast amount of dark frames for model training, and fixed in the inference stage~\cite{feng2023learnability}. However, there also exist some scenarios where the working environment differs significantly from the calibration one, such as a heated sensor under high-speed continuous shooting or relatively long exposure. In such cases, the actual dark shadings in inference may deviate from the pre-calibrated one obviously, and an online re-calibration is plausible. Given that the available time for re-calibration is often constrained, the denoising model needs to perform well even with dark shadings calibrated from a limited number of dark frames.

We employ the LRID dataset to study the aforementioned issue for its separated hot and non-hot dark frames, which perfectly echoes the discussed scenario of varying dark shadings. With the denoising model trained in \Sref{subsec:denoise_performance}, we test its performance \wrt dark shadings calibrated with varying numbers of dark frames. As reported in \Tref{tab:ablation_study_n_ds}, our proposal is visibly robust to dark shading accuracy, evidenced by the fact that the overall PSNR value would only drop $0.1\mathrm{dB}$ when using just $10$ dark frames, which is a reasonable setup for inference-phase dark shading re-calibration. Moreover, a cross-comparison between Tables~\ref{tab:lrid_res} and~\ref{tab:ablation_study_n_ds} shows that, even with only $2$ dark frames for re-calibration, our method can still outperform all existing self-supervised approaches.

\begin{table}[t]
\centering
\caption{Denoising \wrt $\#$dark frames used in inference-phase online dark shading calibration (PSNR/SSIM).}
\begin{adjustbox}{max width=\linewidth}
\begin{NiceTabular}{c|cccccccc}
\toprule
    \textbf{\makecell{$\#$Dark\\frame}} & \textbf{1}  & \textbf{2} & \textbf{5} & \textbf{10} & \textbf{20} & \textbf{50} & \textbf{100} & \textbf{180} \\
\midrule
    \multirow{2}*{\textbf{\makecell{Overall}}}  
    & 44.34 & 44.67 & 44.88 & 44.98 & 45.04 & 45.06 & 45.07 & 45.08 \\
    & 0.9720 & 0.9746 & 0.9758 & 0.9767 & 0.9775 & 0.9778 & 0.9777 & 0.9779 \\
\bottomrule
\end{NiceTabular}
\end{adjustbox}
\label{tab:ablation_study_n_ds}
\end{table}

\paragraph{Data efficiency of our noise synthesis pipeline}
While dark frames can be easily collected, it would still be helpful to decrease its total required number for further cost reduction (\eg, storing dark frames of a high-resolution sensor may consume several terabytes of space). Therefore, we here examine the denoising performance when fewer dark frames are available for training. 

We use the SID and ELD datasets and train denoising networks with varying numbers of dark frames. The results are shown in \Tref{tab:ablation_data_efficiency}. Our proposed noise synthesis pipeline exhibits strong data efficiency, with only $\sim 0.1\mathrm{dB}$ PSNR decrease when just $10$ dark frames are available for each ISO. We attribute this high data efficiency to the fact that the total number of synthesized noisy images is proportional to the number of dark frames. For example, pairing between $10$ clean images and $10$ dark frames leads to $100$ distinct noisy-clean image pairs.

\begin{table}[t]
\centering
\caption{Data efficiency of our noise synthesis pipeline \wrt $\#$dark frames per ISO used in training (PSNR/SSIM).}
\begin{adjustbox}{max width=\linewidth}
\begin{NiceTabular}{c|ccccc}
\toprule
    \textbf{$\#$Dark frames per ISO} & \textbf{400} & \textbf{280} & \textbf{160} & \textbf{40} & \textbf{10} \\
\midrule
    \multirow{2}*{\textbf{\makecell{SID}}}
    & 40.85 & 40.93 & 40.91 & 40.83 & 40.80 \\
    & 0.9478 & 0.9485 & 0.9481 & 0.9453 & 0.9445 \\
\midrule
    \multirow{2}*{\textbf{\makecell{ELD}}} 
    & 46.43 & 46.35 & 46.42 & 46.33 & 46.31 \\
    & 0.9834 & 0.9829 & 0.9835 & 0.9827 & 0.9828 \\
\bottomrule
\end{NiceTabular}
\end{adjustbox}
\label{tab:ablation_data_efficiency}
\end{table}

\subsection{Analyses of signal-independent noise synthesis}\label{subsec:analyze_signal_ind_noise}
We provide detailed analyses regarding explicit signal-independent noise profiling and the high-bit-depth noise recovery strategy, wishing to prompt practically simple yet realistic noise synthesis in future research. 

\paragraph{Statistical signal-independent noise profiling}
Statistical signal-independent noise profiling and synthesis work by fitting a statistical model to (decomposed) dark frames, and resampling from it. Despite the continual emergence of new methods with increasingly accurate (and complex) models, there is a lack of comprehensive justification regarding the inherent limitations of statistical signal-independent noise profiling compared to direct sensor sampling, motivating our following discussion. Specifically, since the statistical models are fit on dark frames, noise resampled from them essentially resembles new dark frames, which can neither achieve the same level of realism nor characterize the real-world diversity compared to those captured directly by sensors. For a thorough demonstration, we experiment to profile signal-independent noise as follows:
\begin{itemize}
    \item \textbf{Tukey-Lambda distribution:} As studied in ELD~\cite{wei2021physics}, signal-independent noise appears to be long-tailed, and hence can be reasonably characterized with the Tukey-Lambda distribution.
    \item \textbf{Gaussian mixture model (GMM):} While not employed in any existing methods to the best of our knowledge, we also explore GMM in noise profiling owing to its extensive modeling ability.
\end{itemize}

We conduct experiments on dark frames of the Sony-A7S2 camera. For GMM, we empirically set the number of Gaussian components to $100$ while allowing non-diagonal covariance matrices. To ensure optimal performance, we randomly select $50$ frames, center-crop them to $512\times512$ patches, and perform noise disentanglement similar to PNNP~\cite{feng2023physics} (\ie, subtracting dark shadings and modeling banding noise separately) to extract the i.i.d. components for distribution fitting. With the fitted distributions, we perform quantile–quantile tests \wrt both the training data to assess modeling accuracy, and with unseen dark frames to evaluate generalizability. We also employ them as alternatives to direct sampling for training denoising networks to show their ability comprehensively. 

\begin{table}[t]
\centering
\begin{adjustbox}{max width=\linewidth}
\addtolength{\tabcolsep}{-1em}
\begin{NiceTabular}{cc}
    \includegraphics[width=\linewidth]{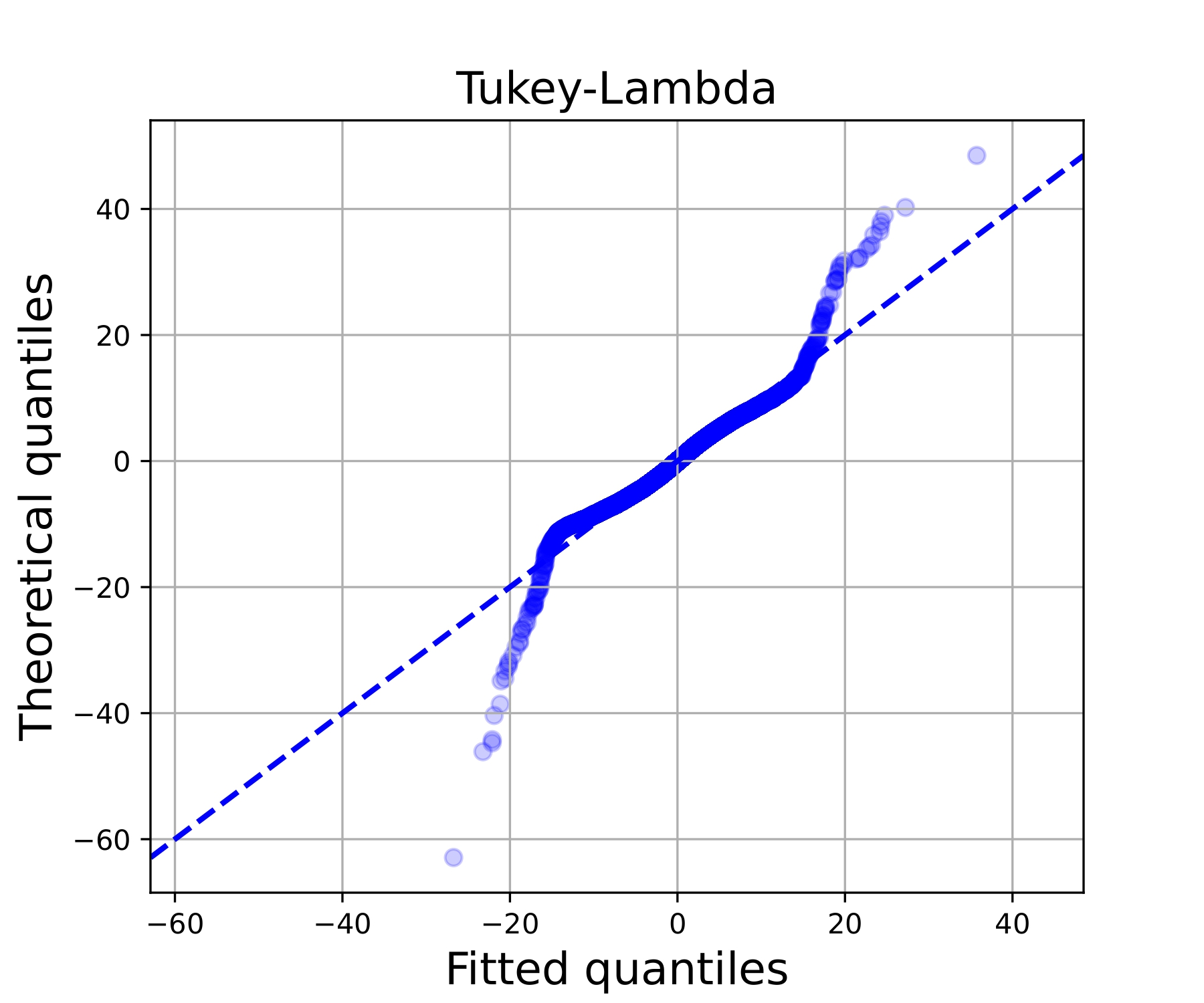} & \includegraphics[width=\linewidth]{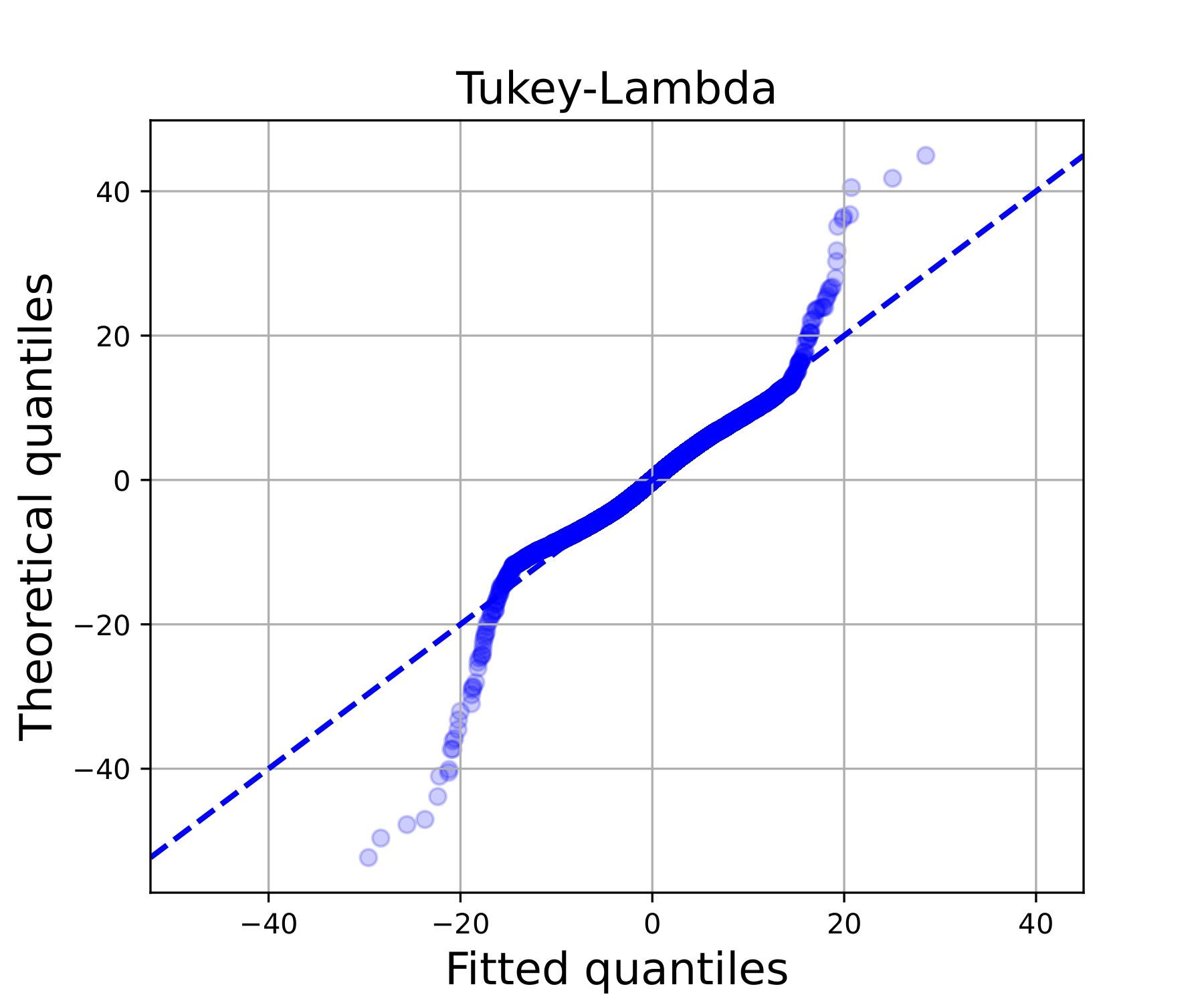} \\
    
    \includegraphics[width=\linewidth]{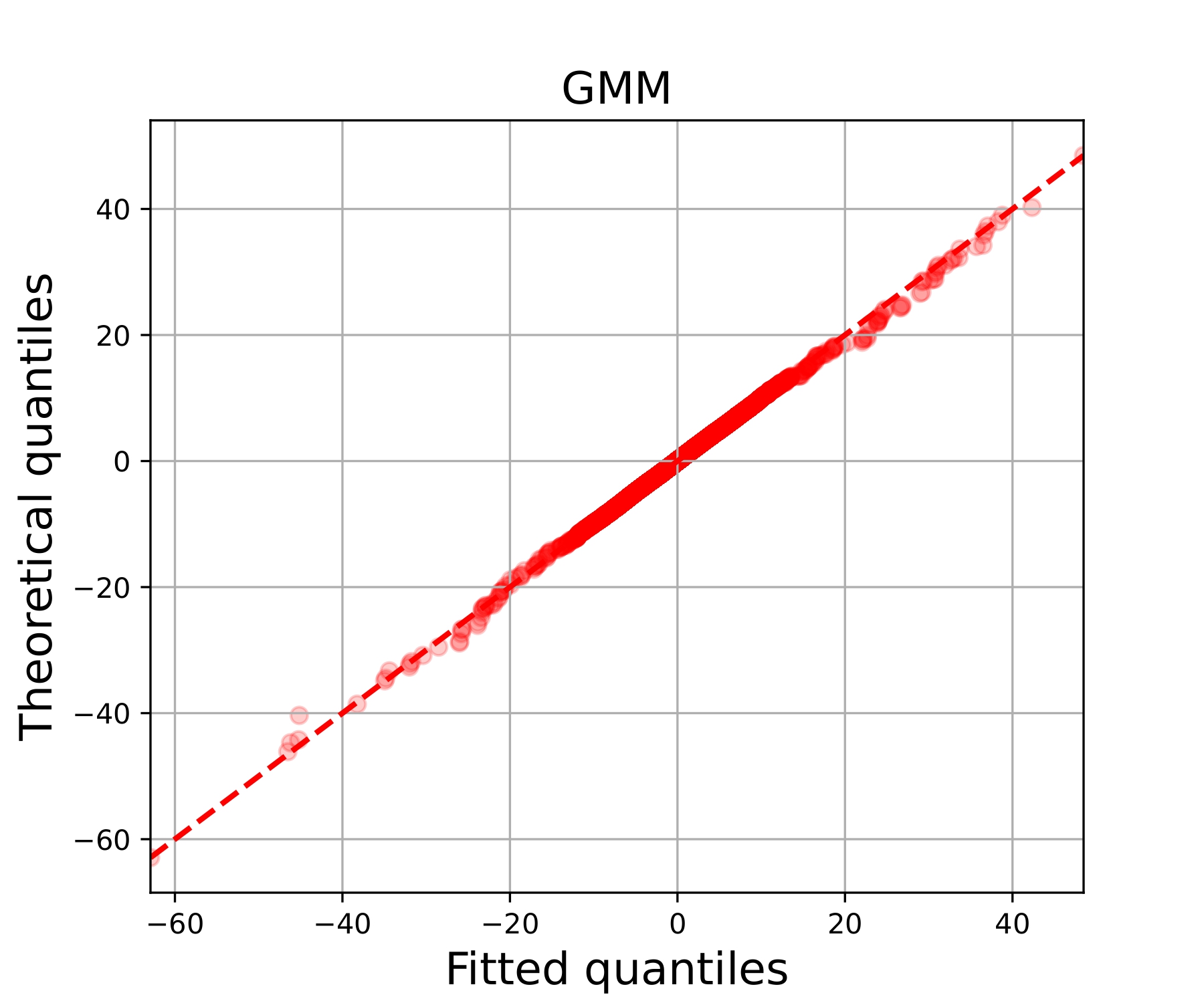} & \includegraphics[width=\linewidth]{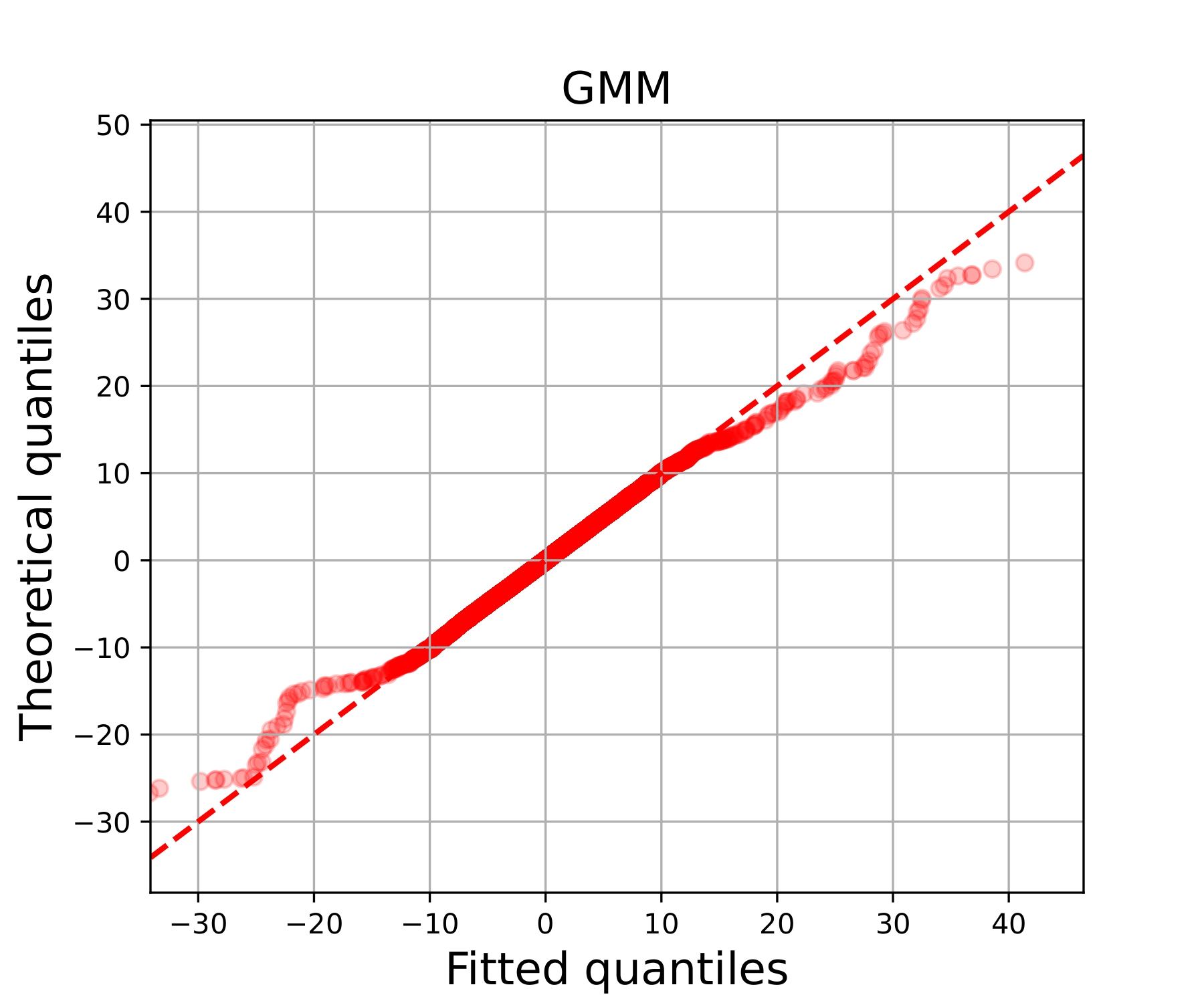} \\
\end{NiceTabular}
\end{adjustbox}
\captionof{figure}{Quantile-Quantile plots of different models on fitting the signal-independent noise. Left: Resampled data versus data used for model fitting (\ie, profiling accuracy). Right: Resampled data versus data from unseen dark frames (\ie, generalization accuracy).}
\label{fig:para_modeling}
\end{table}

\begin{table}[t]
\centering
\caption{Denoising performance with different signal-independent noise profiling methods (PSNR/SSIM).}
\begin{adjustbox}{max width=\linewidth}
\begin{NiceTabular}{c|ccc}
\toprule
    \textbf{Dataset} & \textbf{Tukey-Lambda} & \textbf{GMM} & \textbf{Direct sampling} \\
\midrule
    \makecell{\textbf{SID}} & \makecell{40.56 / 0.9445} & \makecell{40.54 / 0.9454} & \makecell{40.85 / 0.9478} \\
\midrule
    \makecell{\textbf{ELD}} & \makecell{45.63 / 0.9753} & \makecell{45.92 / 0.9815} & \makecell{46.43 / 0.9834} \\
\bottomrule
\end{NiceTabular}
\end{adjustbox}
\label{tab:explicit_noise_model}
\end{table}

\Fref{fig:para_modeling} illustrates the quantile-quantile results. Tukey-Lambda distribution, as a simple model, fails to fully capture the true noise distribution during training, resulting in inaccurate noise samples. While GMM can perfectly fit the training data, its resampling results still exhibit deviations compared to unseen dark frames. We attribute this phenomenon to the complexity of the underlying noise distribution, which cannot be fully characterized by the dark points used in GMM fitting. \Tref{tab:explicit_noise_model} reports the denoising performance, which aligns with the quantile-quantile studies.

\paragraph{Effect of HBNR}
Given our exclusion of HBNR as mentioned in \Sref{subsec:signal_independent_noise}, and its demonstrated importance in terms of P-G, ELD, and SFRN-based signal-independent noise synthesis~\cite{zhang2021rethinking}, we empirically study the denoising performance with and without HBNR in the context of our noise synthesis pipeline. For comprehensiveness, we employ all three aforementioned datasets for this ablation study. \Tref{tab:ablation_hbnr} records the results. Evidenced by the comparable quantitative performance (\ie, less than $0.1\mathrm{dB}$ difference in PSNR) among all the datasets, we can conclude that the high-bit noise recovery strategy does not demonstrate any obvious benefits in our proposal.

\begin{table}[t]
\centering
\caption{Effect of HBNR in the context of our proposed noise synthesis pipeline (PSNR/SSIM).}
\begin{adjustbox}{max width=\linewidth}
\begin{NiceTabular}{cc|cc|cc}
\toprule
    \multicolumn{2}{c}{\textbf{SID}} & \multicolumn{2}{c}{\textbf{ELD}} &\multicolumn{2}{c}{\textbf{LRID}} \\
\midrule
    \textbf{\makecell{w/ HBNR}} & \textbf{\makecell{w/o HBNR}} & \textbf{\makecell{w/ HBNR}} & \textbf{\makecell{w/o HBNR}} & \textbf{\makecell{w/ HBNR}} & \textbf{\makecell{w/o HBNR}} \\
\midrule
    40.91 & 40.85 & 46.38 & 46.43 & 45.03 & 45.08 \\
    0.9485 & 0.9478 & 0.9833 & 0.9834 & 0.9776 & 0.9779 \\
\bottomrule
\end{NiceTabular}
\end{adjustbox}
\label{tab:ablation_hbnr}
\end{table}

\section{Conclusion}\label{sec:conclusion}
This paper introduces a practically simple noise synthesis pipeline for self-supervised RAW image denoising. While delivering state-of-the-art denoising performance, our proposal eliminates the labor-intensive system gain calibration, the parametric signal-independent noise profiling, and the high-bit-depth noisy recovery processes, which are commonly required in existing methods. 

A practical limitation of our approach lies in the need for known AG. While AG is typically accessible to direct sensor users such as camera manufacturers or end users of most industrial cameras, end users of consumer cameras and smartphones are often provided ISO values only with the AG hidden inside, making hypothesizing the system gain challenging. In such a case, the base ISO may provide helpful hints for linking ISO and AG.

For future work, we plan to incorporate dark frames from multiple sensor models to investigate sensor-agnostic denoising. Another potential direction lies in applying the current pipeline to other types of sensors, such as EMCCD.

{
    \small
    \bibliographystyle{ieeetr}
    \bibliography{main}
}

\end{document}


\maketitle

\section{Quantum efficiency of modern imaging sensors}

Here, we present the typical fill-factor-included spectral-averaging quantum efficiency (QE) of modern imaging sensors. Specifically, we plot the QE values of 393 different camera models sourced from DxOMark\footnote{\url{https://www.dxomark.com/}} and summarized by Photons to Photos\footnote{\url{https://www.photonstophotos.net/Charts/Sensor_Characteristics.htm}}. As illustrated in \fref{fig:quantum_efficiency}, a dominant portion of sensors exhibits QE values in the range of $\left(30\%, 70\%\right)$. Another noteworthy observation is that sensor models with QE values lower than $30\%$ were typically released before (or around) the year $2010$. This timestamp marks a transition from front-illuminated CMOS sensors to back-illuminated counterparts, leading to a substantial increase in QE. To conclude, the analysis of QE reinforces our hypothesis-based shot noise synthesis method as being generally applicable, particularly for modern sensors.

\begin{table*}[h]
\centering
\begin{adjustbox}{max width=\linewidth}
\addtolength{\tabcolsep}{-0.4em}
\begin{NiceTabular}{cccc}
    \includegraphics[width=0.5\textwidth]{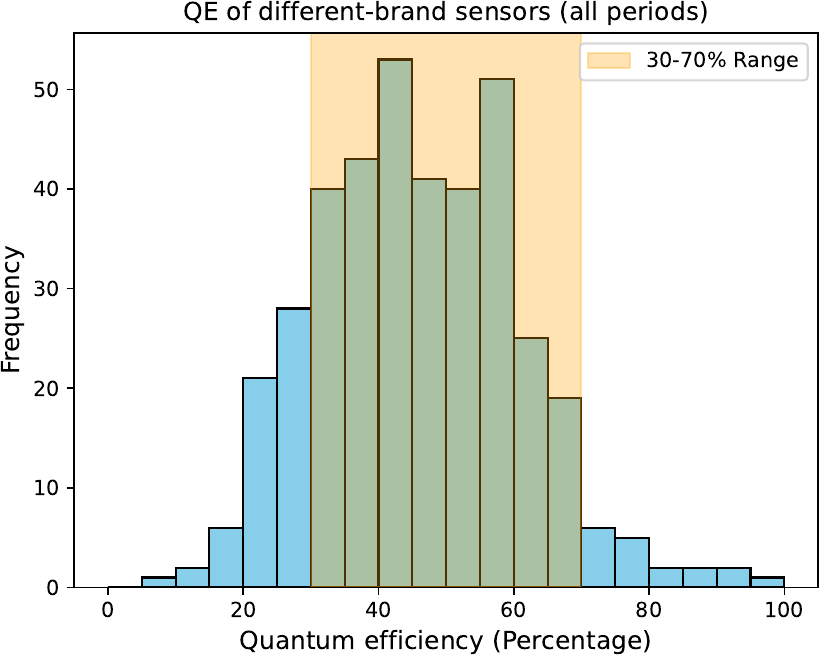} & 
    \includegraphics[width=0.5\textwidth]{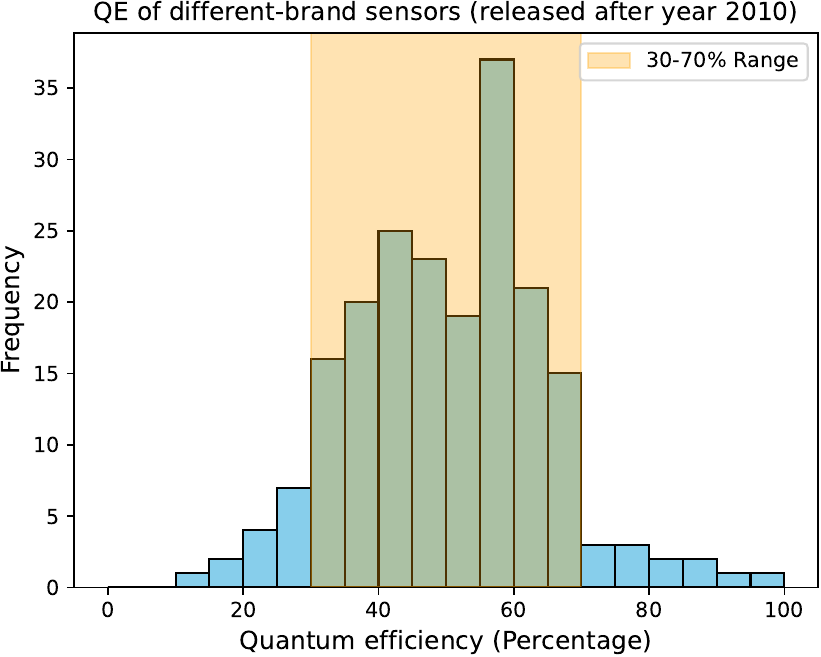} \\
\end{NiceTabular}
\end{adjustbox}
\captionof{figure}{Quantum efficiency statistics of different-brand imaging sensors. There are $393$ data points in total, with $204$ representing models released later than the year 2010. Although some data points appear to be outliers (\eg, there unlikely exists a CMOS sensor with a $100\%$ QE), we keep them in the plots to stick to the original data.}
\label{fig:quantum_efficiency}
\end{table*}


\section{Importance of different noise components}
Here, we study the importance of different components in noise synthesis for training the denoising network. We use the SID and ELD datasets for experimental setup and disable different noise components during noise synthesis. As summarized in \Tref{tab:noise_importance}, neither shot nor signal-independent noise can be omitted from the noise synthesis pipeline, otherwise, the denoising performance would diminish significantly. Also, dark shading plays an important role in improving the denoising accuracy.

\begin{table}[t]
    \centering
    \caption{Denoising performance \wrt different noise components used during synthesis.}
    \begin{adjustbox}{max width=\linewidth}
    \begin{NiceTabular}{ccc|cc|cc}
    \toprule
    \multicolumn{3}{c}{\textbf{Noise component used in synthesis}} & 
    \multicolumn{2}{c}{\textbf{SID}} & 
    \multicolumn{2}{c}{\textbf{ELD}} \\
    \midrule
    Photon shot noise & Sampled dark frame & Dark shading correction & PSNR & SSIM & PSNR & SSIM \\
    \midrule
    $\checkmark$ & & & 39.81 & 0.9182 & 44.95 & 0.9613 \\
    & $\checkmark$ & & 39.95 & 0.9412 & 44.90 & 0.9708 \\
    \midrule
    $\checkmark$ & $\checkmark$ & & 40.14 & 0.9438 & 45.49 & 0.9786 \\
    & $\checkmark$ & $\checkmark$ & 40.37 & 0.9436 & 45.60 & 0.9788 \\
    $\checkmark$ & & $\checkmark$ & 38.12 & 0.8922 & 44.16 & 0.9673 \\
    \midrule
    $\checkmark$ & $\checkmark$ & $\checkmark$ & 40.85 & 0.9478 & 46.43 & 0.9834 \\
    \bottomrule
    \end{NiceTabular}
    \end{adjustbox}
    \label{tab:noise_importance}
    \end{table}


\section{More denoising results}
We present more qualitative denoising results achieved with our noise synthesis pipeline. As illustrated in \fref{fig:denoise_results}, our proposal can help the denoising network effectively smooth real-world noisy images in various scenarios without losing many high-frequency details.

\begin{table*}[t]
\centering
\begin{adjustbox}{max width=\linewidth}
\addtolength{\tabcolsep}{-0.4em}
\begin{NiceTabular}{ccccc}
    & Input & Prediction & Input & Prediction \\
    \multirow{2}{*}{\rotatebox[origin=c]{90}{SID dataset~\cite{chen2018learning}}} &
    \includegraphics[width=0.25\textwidth]{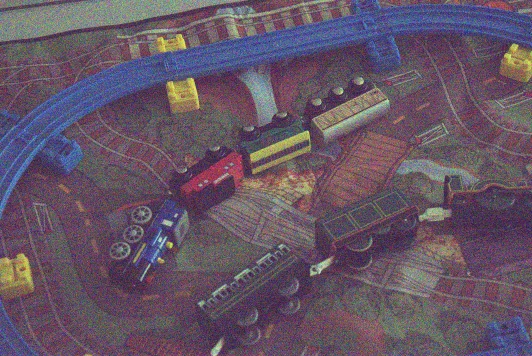} & 
    \includegraphics[width=0.25\textwidth]{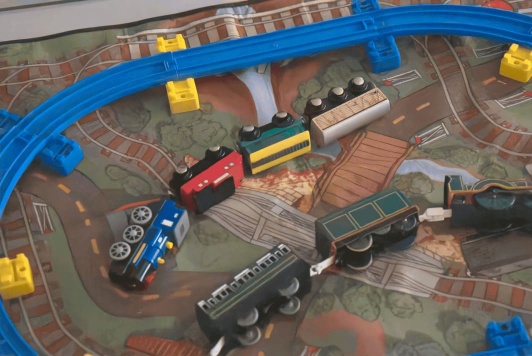} & 
    \includegraphics[width=0.25\textwidth]{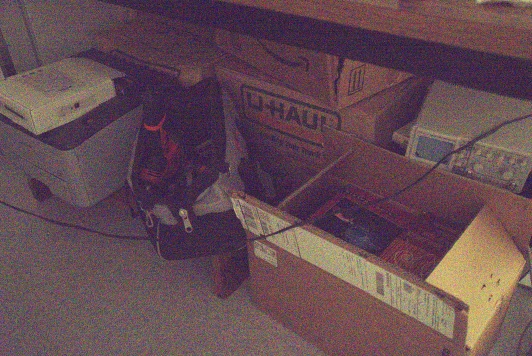} &
    \includegraphics[width=0.25\textwidth]{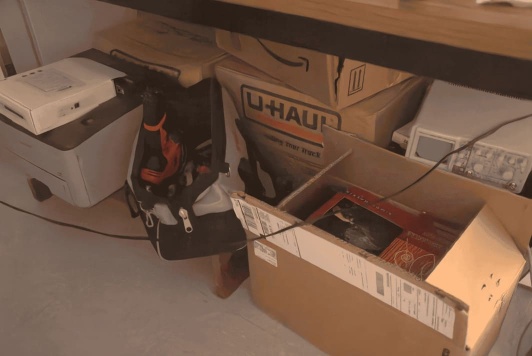} \\ 
    & 
    \includegraphics[width=0.25\textwidth]{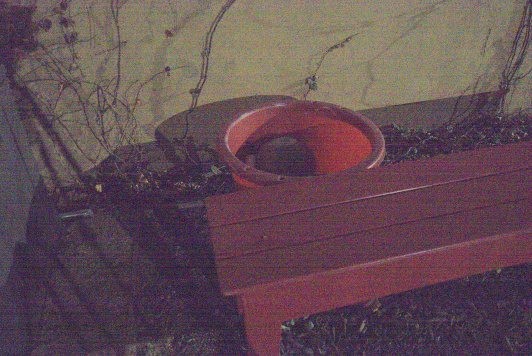} & 
    \includegraphics[width=0.25\textwidth]{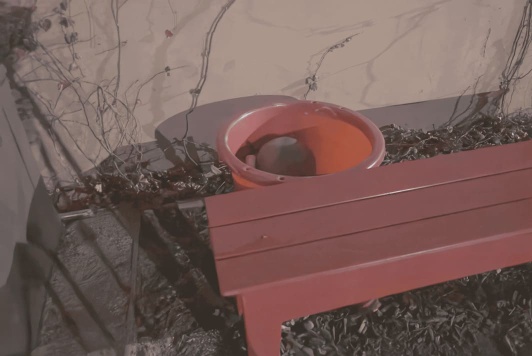} & 
    \includegraphics[width=0.25\textwidth]{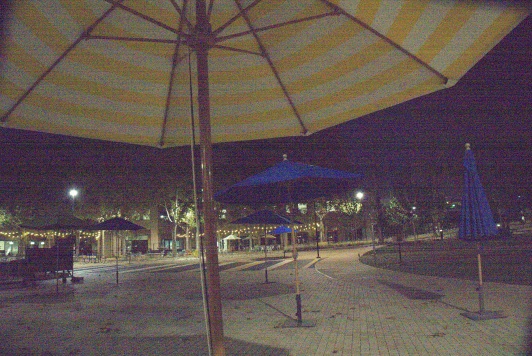} &
    \includegraphics[width=0.25\textwidth]{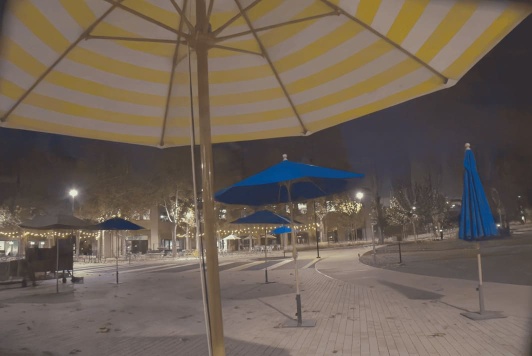} \\ 
    
    \multirow{2}{*}{\rotatebox[origin=c]{90}{ELD dataset~\cite{wei2021physics}}} &
    \includegraphics[width=0.25\textwidth]{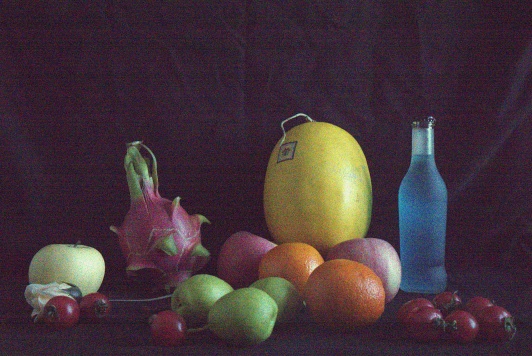} & 
    \includegraphics[width=0.25\textwidth]{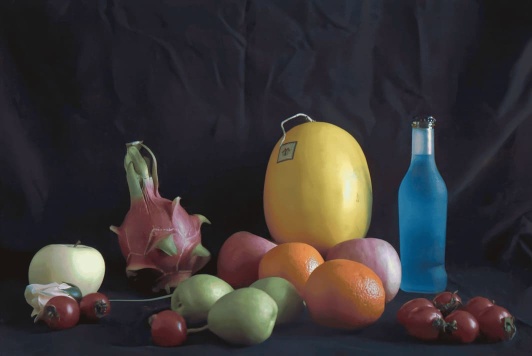} & 
    \includegraphics[width=0.25\textwidth]{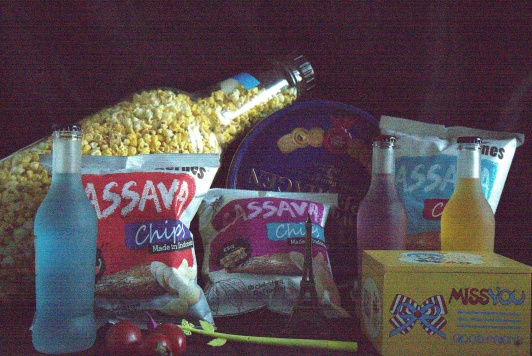} &
    \includegraphics[width=0.25\textwidth]{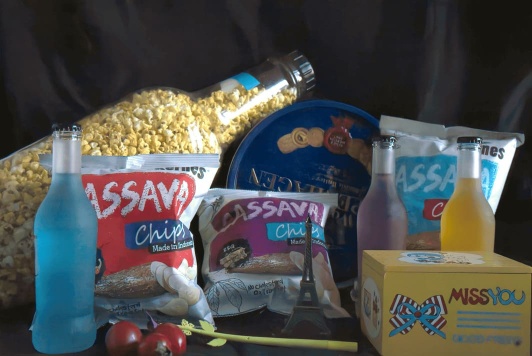} \\ 
    & 
    \includegraphics[width=0.25\textwidth]{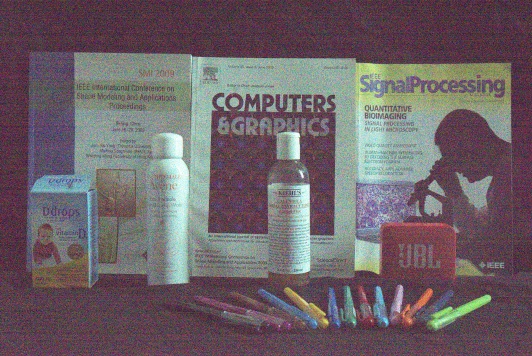} & 
    \includegraphics[width=0.25\textwidth]{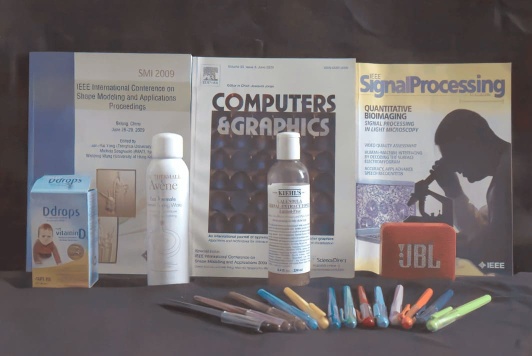} & 
    \includegraphics[width=0.25\textwidth]{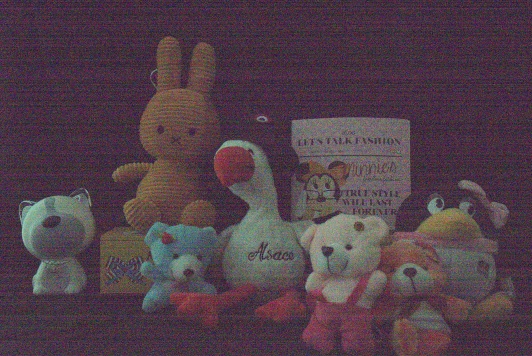} &
    \includegraphics[width=0.25\textwidth]{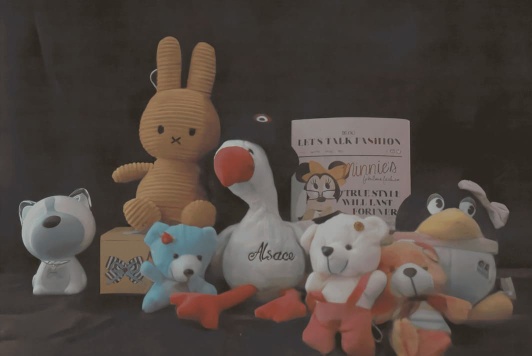} \\ 
    
    \multirow{2}{*}{\rotatebox[origin=c]{90}{LRID dataset~\cite{feng2023learnability}}} &
    \includegraphics[width=0.25\textwidth]{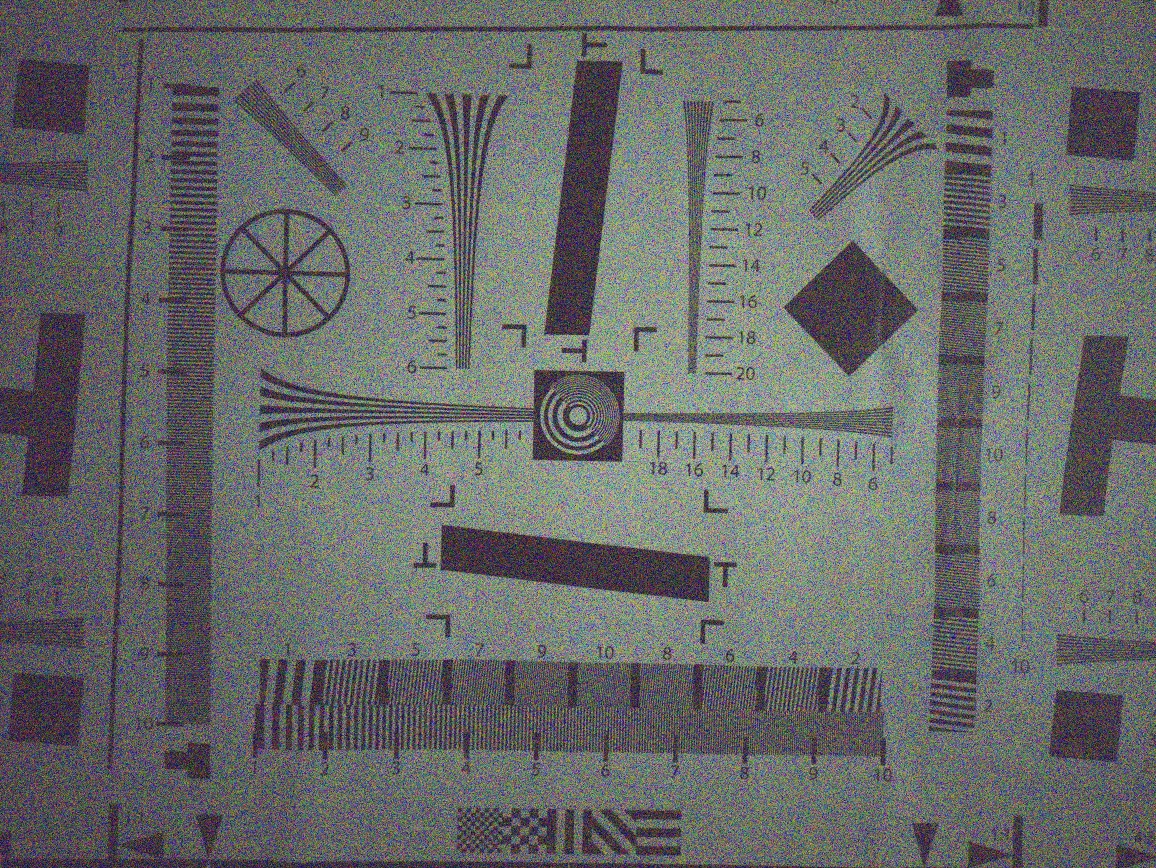} & 
    \includegraphics[width=0.25\textwidth]{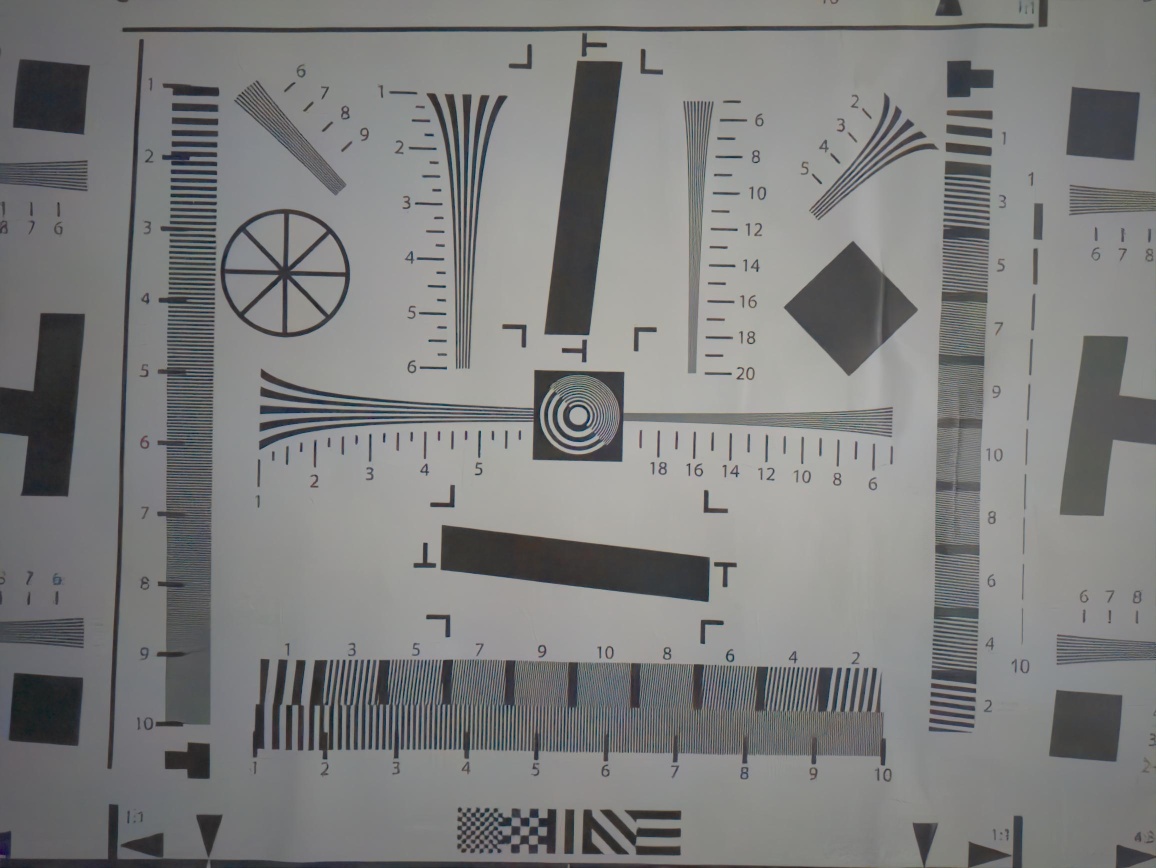} & 
    \includegraphics[width=0.25\textwidth]{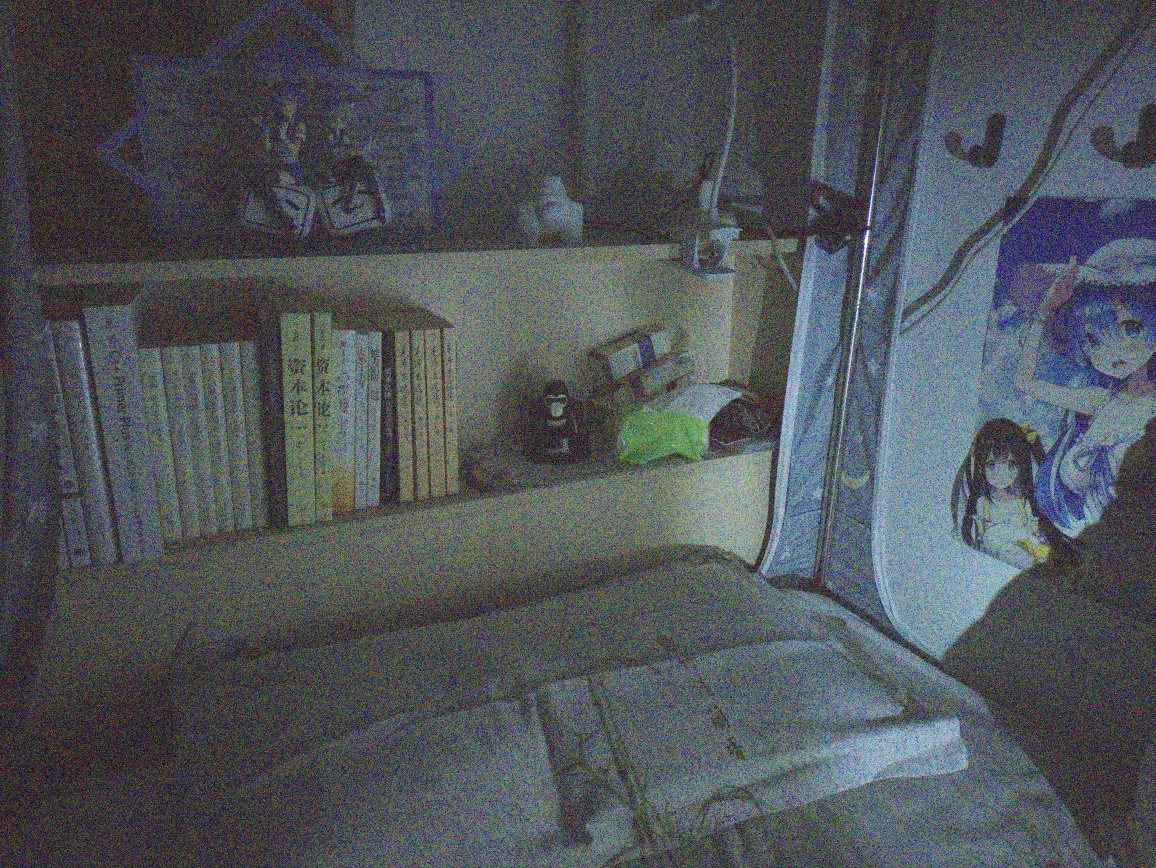} &
    \includegraphics[width=0.25\textwidth]{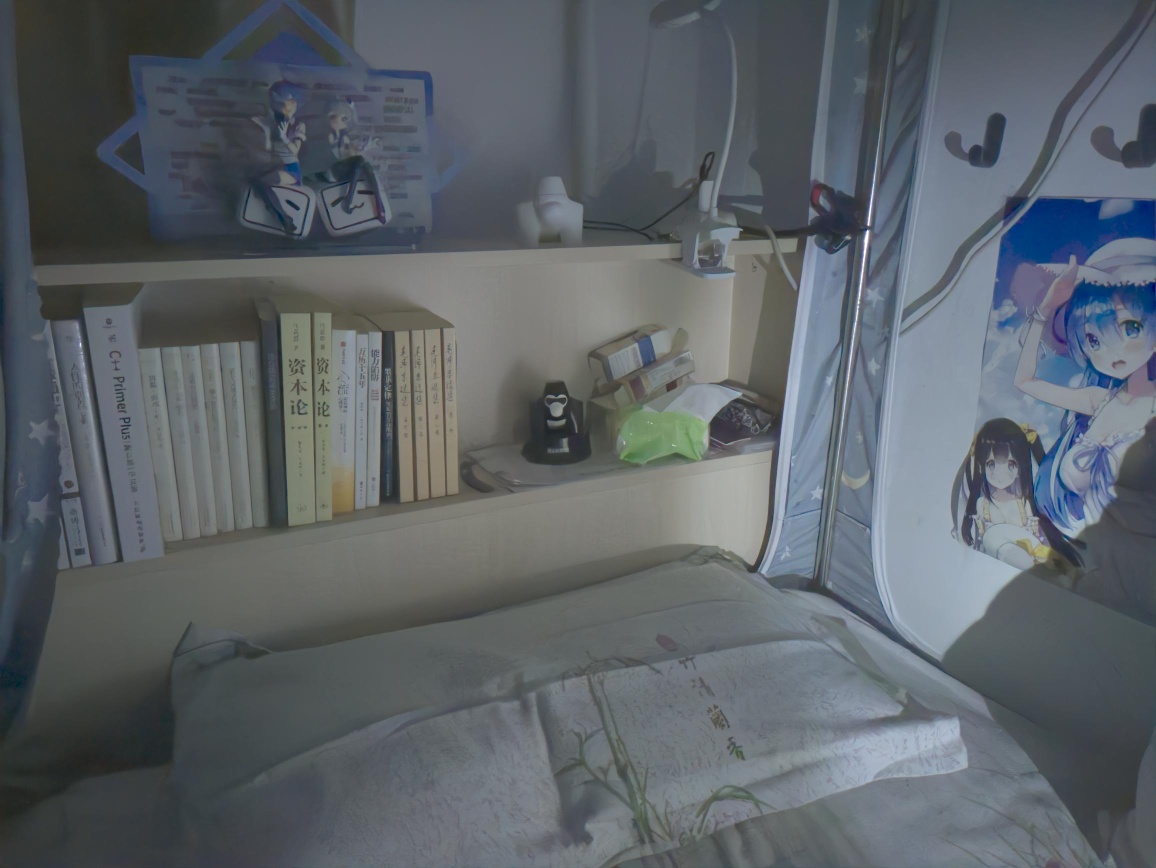} \\ 
    & 
    \includegraphics[width=0.25\textwidth]{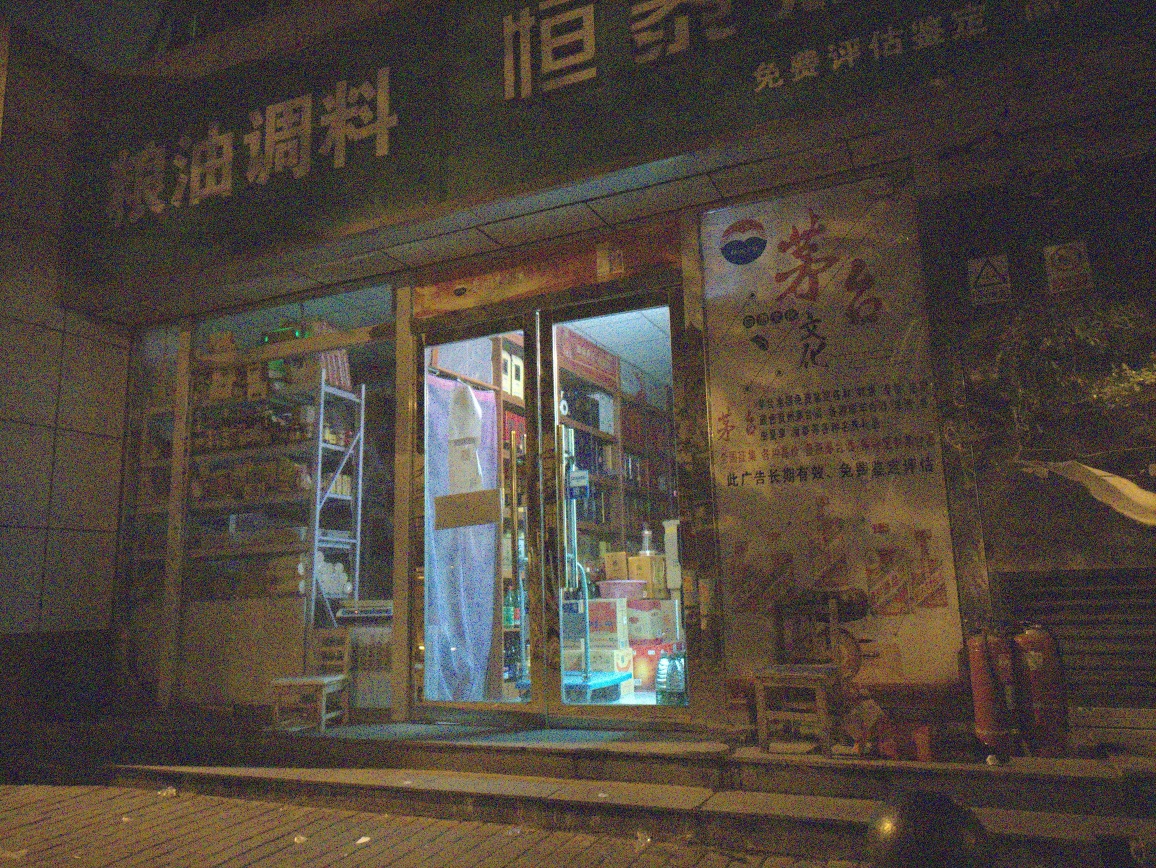} & 
    \includegraphics[width=0.25\textwidth]{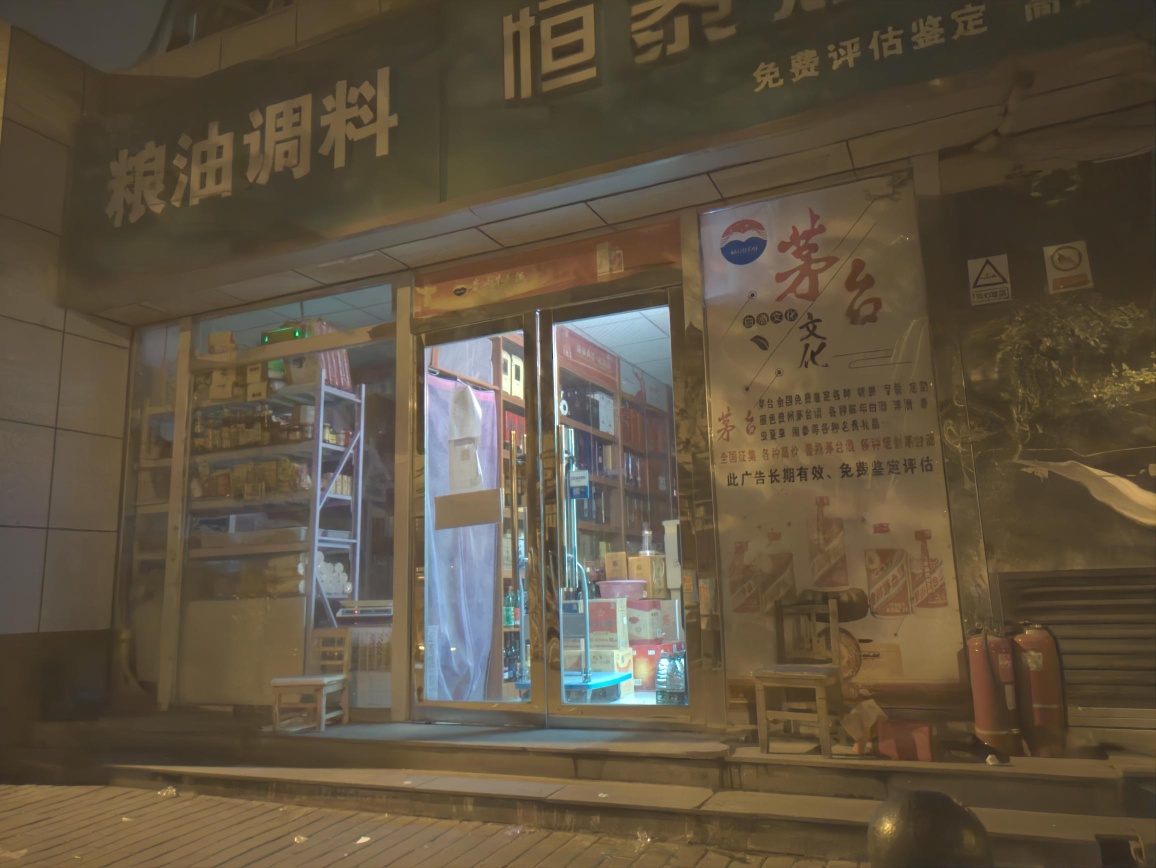} & 
    \includegraphics[width=0.25\textwidth]{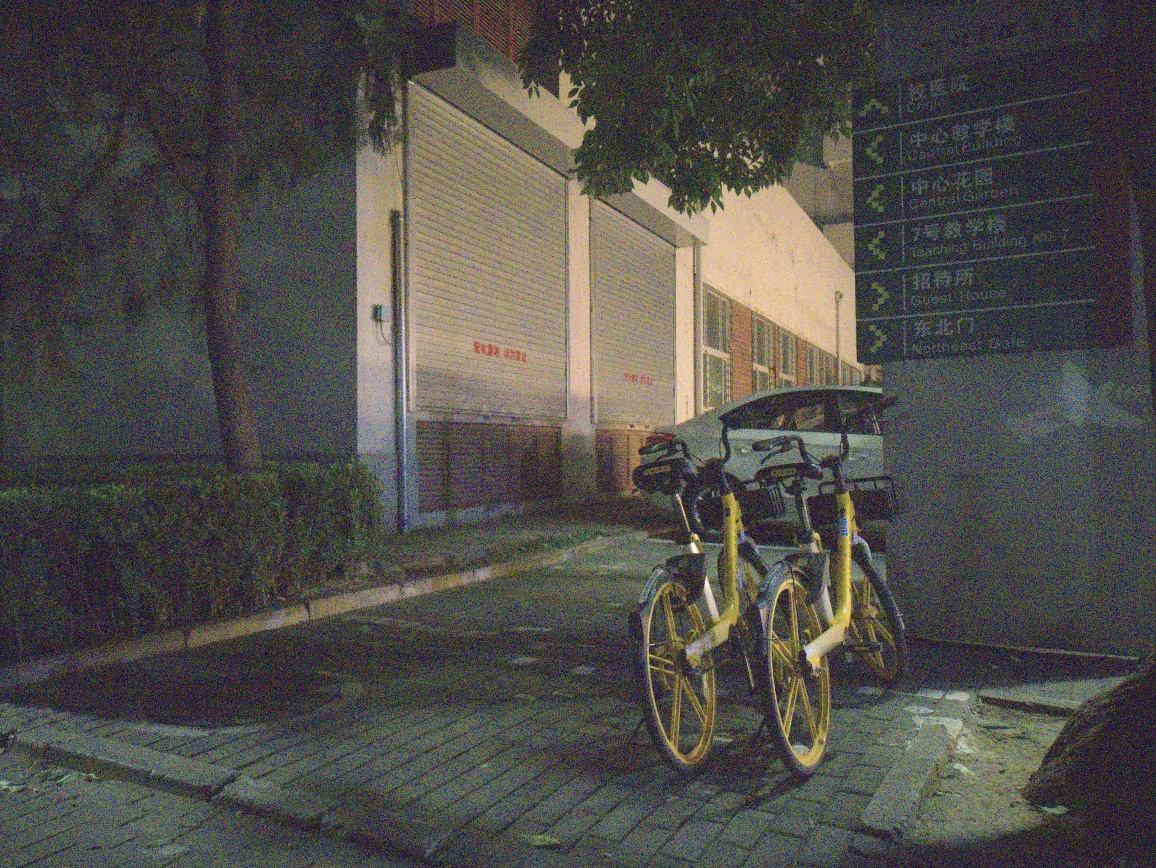} &
    \includegraphics[width=0.25\textwidth]{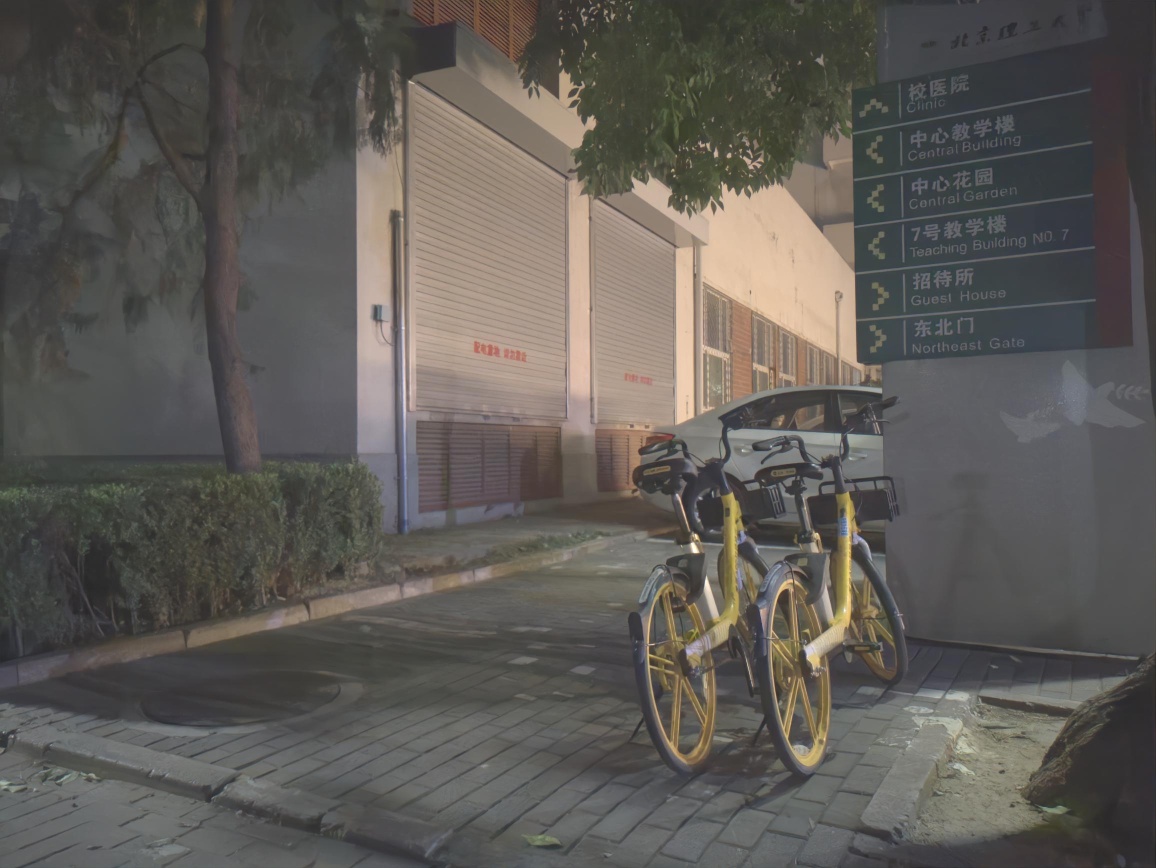} \\ 

\end{NiceTabular}
\end{adjustbox}
\captionof{figure}{Qualitative denoising results achieved by our proposed method on distinct datasets. All images are brightness-adjusted via a digital gain, and gamma-corrected for better visualization.}
\label{fig:denoise_results}
\end{table*}

{
    \small
    \bibliographystyle{ieeetr}
    \bibliography{main}
}